\documentclass[journal]{vgtc-no-name}              

\vgtccategory{Research}
\vgtcpapertype{system}
\graphicspath{{./figs/}{./supplement-figs/}} 
\usepackage{graphicx}
\usepackage[svgnames]{xcolor}
\usepackage{tabularx}
\usepackage{colortbl}
\usepackage{array}
\usepackage{amssymb}
\usepackage{caption}
\usepackage{wrapfig}
\newcommand {\mm}[1] {\ifmmode{#1}\else{\mbox{\(#1\)}}\fi}

\newcommand{\Rspace}        {\mm{\mathbb{R}}}

\newcommand{\Xspace}        {\mm{\mathbb{X}}}

\newcommand{\Ucal}        {\mm{\mathcal U}}
\newcommand{\Vcal}        {\mm{\mathcal V}}
\newcommand{\Wcal}        {\mm{\mathcal W}}

\newcommand{\tool}{\textsl{Explainable Mapper}}

\usepackage{xcolor,graphicx}
\newcommand{\mycolorbox}[2]{{\setlength{\fboxsep}{1pt} \colorbox{#1}{#2}}}

\newcommand{\para}[1]{\vspace{0mm}\noindent{\textbf{#1}}}

\definecolor{lightblue}{RGB}{220, 235, 250}  
\definecolor{darkblue}{RGB}{150, 180, 210}  
\definecolor{pastelpurple}{RGB}{235, 220, 250} 
\definecolor{darkpurple}{RGB}{180, 150, 210}  
\definecolor{lightorange}{RGB}{250, 230, 210} 
\definecolor{darkorange}{RGB}{210, 160, 120}  
\definecolor{lightarmygreen}{RGB}{220, 230, 210}  
\definecolor{darkarmygreen}{RGB}{150, 160, 140}  
\definecolor{lightpastelyellow}{RGB}{255, 250, 205}
\definecolor{darkpastelyellow}{RGB}{200, 190, 140} 
\definecolor{lightgrey}{RGB}{230, 230, 230} 
\definecolor{darkgrey}{RGB}{160, 160, 160}  

\title{\textbf{{\tool}}: Charting LLM Embedding Spaces 
Using Perturbation-Based Explanation and Verification Agents}
\author{
  Xinyuan Yan, 
  Rita Sevastjanova,
  Sinie van der Ben, 
  Mennatallah El-Assady, and 
  Bei Wang
}
\authorfooter{
\item Xinyuan Yan and Bei Wang are with the University of Utah.
  	E-mails: xinyuan.yan@sci.utah.edu, beiwang@sci.utah.edu. 
\item Rita Sevastjanova, Sinie van der Ben, and Mennatallah El-Assady are with ETH Zürich. 
  	E-mails: rita.sevastjanova@inf.ethz.ch, vandersi@ethz.ch, menna.elassady@ai.ethz.ch. 
\item The first two authors contributed equally to this work.  
}

\abstract{
Large language models (LLMs) produce high-dimensional embeddings that capture rich semantic and syntactic relationships between words, sentences, and concepts. Investigating the topological structures of LLM embedding spaces via mapper graphs enables us to understand their underlying structures. Specifically, a mapper graph summarizes the topological structure of the embedding space, where each node represents a topological neighborhood (containing a cluster of embeddings), and an edge connects two nodes if their corresponding neighborhoods overlap. However, manually exploring these embedding spaces to uncover encoded linguistic properties requires considerable human effort. To address this challenge, we introduce a framework for semi-automatic annotation of these embedding properties. To organize the exploration process, we first define a taxonomy of explorable elements within a mapper graph such as nodes, edges, paths, components, and trajectories. The annotation of these elements is executed through two types of customizable LLM-based agents that employ perturbation techniques for scalable and automated analysis. These agents help to explore and explain the characteristics of mapper elements and verify the robustness of the generated explanations. We instantiate the framework within a visual analytics workspace and demonstrate its effectiveness through case studies. In particular, we replicate findings from prior research on BERT’s embedding properties across various layers of its architecture and provide further observations into the linguistic properties of topological neighborhoods.

}


\teaser{
  \centering
  \includegraphics[width=\linewidth, alt={An overview of {\tool}.}]{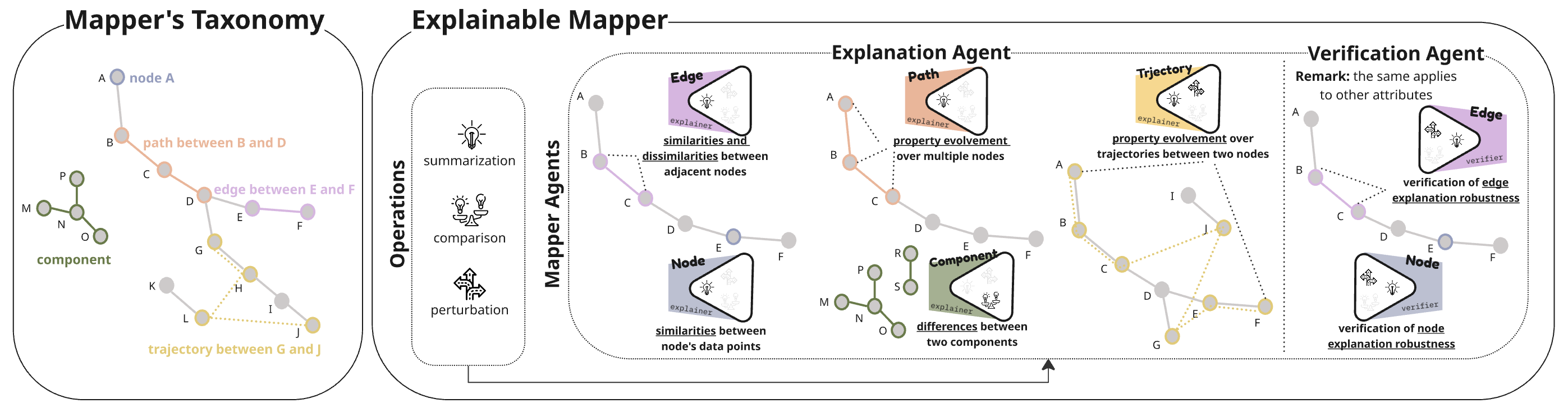}
          \vspace{-17pt}
  \caption{We apply mapper graphs---a widely used tool in topological data analysis and visualization---to investigate the topological structures of LLM embedding spaces. The mapper's taxonomy includes elements such as nodes, edges, paths, components, and trajectories. We introduce the {\tool} workspace and two mapper agents to support embedding investigation. These agents utilize summarization, comparison, and perturbation operations to generate and verify explanations of mapper elements such as the linguistic aspect of clusters, connectivity, and transitions. 
  }
  \label{fig:teaser}
}

\usepackage{microtype}
\begin{document}

\firstsection{Introduction}
\maketitle
Large language models (LLMs) acquire linguistic properties by learning from vast text corpora and have demonstrated remarkable performance across a wide range of natural language processing (NLP) and understanding (NLU) tasks. 
Since their development, researchers have extensively analyzed their learning mechanisms to uncover both their strengths and limitations~\cite{zini2022survey, danilevsky-etal-2020-survey, rogers-etal-2020-primer}.
One key area of investigation has been the study of contextual word embeddings, which dynamically evolve across different layers of the model's architecture~\cite{ethayarajh-2019-contextual,sevastjanova2021explaining}. 
Model developers, computational linguists, and NLP experts are especially interested in analyzing such embedding learning behaviors~\cite{boggust2022embedding}.

A common approach to exploring contextual word embeddings involves computational methods combined with visualizations~\cite{huang2023VA+embeddings-star}. 
One widely used technique is dimensionality reduction, which projects embeddings into a lower-dimensional space while preserving key structural relationships.
These reduced representations are often visualized using scatter plots~\cite{huang2023VA+embeddings-star}, allowing for a closer examination of word neighborhoods~\cite{boggust2022embedding} and their semantic or syntactic similarities~\cite{sevastjanova2022lmfingerprints}. 
Furthermore, they facilitate tracking shifts in word meanings across various contexts~\cite{sevastjanova2025layerflows} and uncovering potential biases in the model's learned representations~\cite{sevastjanova2022adapters,RathoreDevPhillips2023}. 
Beyond traditional projection-based visualizations, researchers have recently leveraged \emph{mapper graphs}~\cite{SinghMemoliCarlsson2007}---a popular tool from topological data analysis and visualization---to investigate the topological structure of word embeddings~\cite{RathoreChalapathiPalande2021,RathoreZhouSrikumar2023}.  
By constructing a graph representation of the embedding space, mapper graphs reveal clusters, transitions, and connectivity patterns~\cite{ZhouChalapathiRathore2021,RathoreChalapathiPalande2021,RathoreZhouSrikumar2023,PurvineBrownJefferson2023,ZhouZhouDing2023}. 
In this paper, we explore mapper graphs to gain insight into the linguistic properties of contextual word embeddings.

To interpret the embedding properties represented in a mapper graph, it is essential to analyze its various structural elements. 
As shown in previous work~\cite{RathoreChalapathiPalande2021,RathoreZhouSrikumar2023}, each element of the mapper graph, such as nodes, edges, and overall connectivity patterns, carries insights into how embeddings are organized and evolve across different contexts. 
For instance, nodes represent topological neighborhoods---clusters of embeddings with shared characteristics---and summarizing their properties has revealed dominant linguistic features, such as common semantic themes or syntactic roles. 
Edges, which indicate overlapping data points between neighboring nodes, help to understand gradual transitions and semantic shifts within the embedding space. 

The explorable space of a mapper graph representing word embeddings is typically vast, as it often visualizes thousands of tokens simultaneously to provide meaningful insights~\cite{RathoreChalapathiPalande2021,RathoreZhouSrikumar2023}.
The insights that can be drawn from the data depend on the dataset's coverage of various linguistic properties and functionalities.
Moreover, as shown by prior work, word embeddings encode multiple linguistic properties simultaneously, some of them being more prominent than others~\cite{rogers-etal-2020-primer}.
As a result, there is usually more than one observation that can summarize the properties of a mapper element. 
Since the embedding properties are not yet fully understood, users typically generate new hypotheses while exploring the embedding space and strive to verify their validity. 
The hypothesis generation process thus can \emph{diverge} to have a large set of possible explanations. At the same time, the verification step aims to \emph{converge} the space to those hypotheses that can be verified.

To assist users in analyzing the structural elements of the mapper graph while initially diverging in the explanation space and thereafter converging the produced explanations to the reliable ones, we introduce the concept of \emph{mapper agents}. 
In particular, we introduce two types of agents, i.e., an \emph{Explanation Agent} that can be used to create explanation candidates of a selected mapper element and a \emph{Verification Agent} that is used to verify the generated explanations against their robustness. 
These two agent classes support distinct operations, i.e., \emph{summarization}, \emph{comparison}, and \emph{perturbation}, to support explanation generation and verification.
In this paper, we propose a set of LLM-based agent instances, i.e., explainers and verifiers, to explain and verify each mapper element. 
We instantiate this concept within a visual analytics workspace called {\tool}, which enables users to explore the embedding space, interact with mapper elements, and produce reliable and robust insights into embedding properties.

To summarize, our contributions are: (1) a mapper agent framework that introduces two types of agents for explanation generation and verification, (2) the instantiation of the framework into an {\tool} workspace, and (3) the evaluation of the proposed approach through replication studies.

\section{Technical Background}
\label{sec:background}
In this section, we review technical background surrounding word embeddings and mapper graphs, which serve as the key ingredients behind {\tool}. 

\subsection{Word Embeddings}
Word embeddings represent words as dense vectors in a continuous vector space, capturing semantic and syntactic relationships through their geometric properties. Before words were represented as embeddings via word embedding model or LLMs, one-hot encoding was often used to transform words into numbers. Unlike one-hot encoding that treats words as independent entities, embeddings that represent words with similar meanings are closer together in a vector space. 

Given a large vocabulary $\Wcal$, an embedding model learns a function $h: \Wcal \to \Rspace^d$ that maps each word in $\Wcal$ to a $d$-dimensional vector. Early models such as Word2Vec \cite{mikolov2013distributed} and GloVe \cite{PenningtonSocherManning2014} were the first in NLP to learn word representations from large text datasets, producing static embeddings where each word has a fixed representation regardless of its context. Their static nature, which does not take context into account, creates the problem of polysemy. BERT (Bidirectional Encoder Representations from Transformers) \cite{DevlinChangLee2019} addressed this limitation by introducing contextual embeddings. These embeddings dynamically adjust the vector for a word based on the context.  

Contextual embeddings improved performance across NLP tasks where semantic understanding is important, such as question answering or named entity recognition. In addition to improving various NLP tasks, the embeddings themselves can reveal new information about the linguistic structures encoded by the language models. Kurita \cite{kurita2019measuring} showed through a case study on gender pronoun resolution, that BERT's performance is influenced by gender stereotypes. 
Furthermore, Jawahar \cite{jawahar-etal-2019-bert} found that different layers of BERT are responsible for the encoding of different types of linguistic information. The lower layers tend to encode surface and lexical features, such as part-of-speech (POS) tags and word order. The middle layers capture syntactic properties while the upper layers encode semantic and contextual information, allowing the formation of complex semantic relations.

\subsection{Taxonomy of Mapper Graphs}
\label{sec:taxonomy}

Given a high-dimensional point cloud $\Xspace$ equipped with a real-valued function $f: \Xspace \to \Rspace$, a (classical) mapper graph~\cite{SinghMemoliCarlsson2007} summarizes the topological structure of $\Xspace$ induced by the function $f$. A node in the mapper graph represents a \emph{topological neighborhood}, whereas an edge connects two nodes if their corresponding neighborhoods overlap. In the context of this paper, $\Xspace$ is a set of word embeddings obtained from an LLM.

\begin{figure}[!ht]
    \centering
    \vspace{-2mm}
    \includegraphics[width=\linewidth]{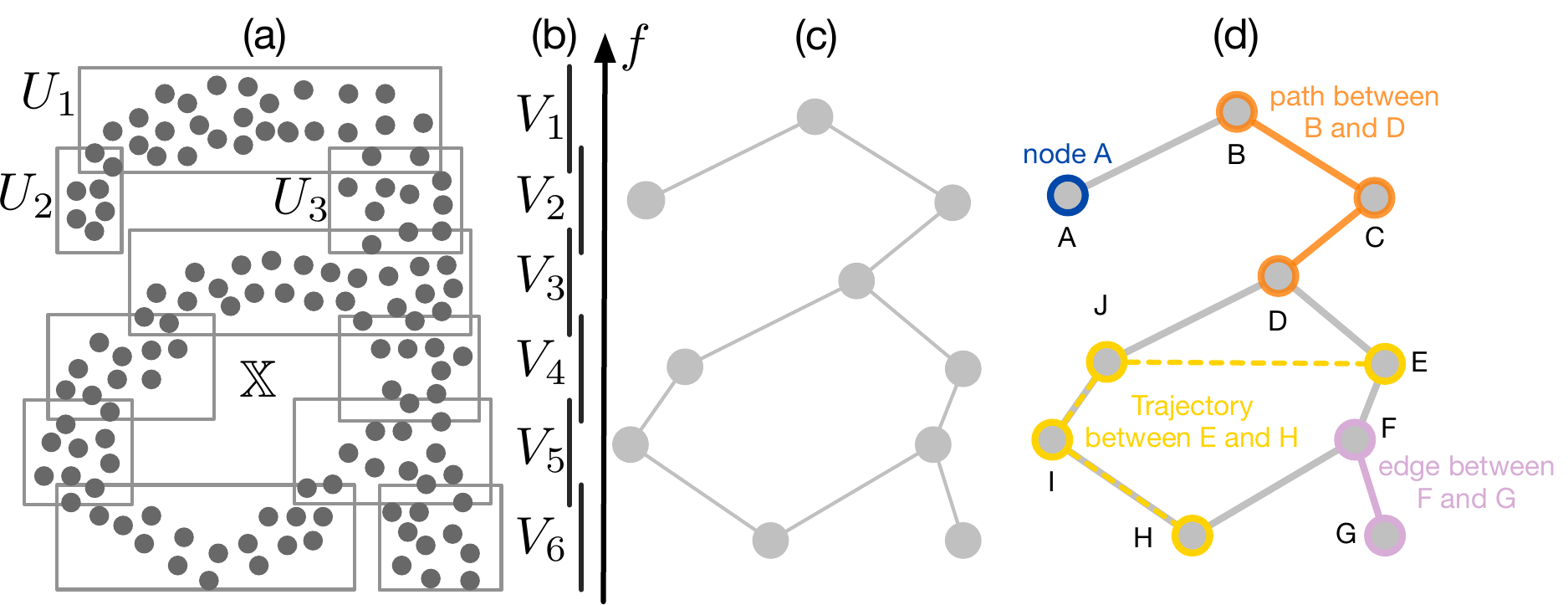}
    \vspace{-6mm}
    \caption{A simple example of computing a mapper graph. 
        A point cloud $\Xspace$ in (a) is equipped with a height function $f$. The cover $\Ucal = \{U_1,U_2,\dots\}$ of $\Xspace$ is induced by a cover $\Vcal = \{V_1, V_2,\dots\}$ of $f(\Xspace)$ that contains 6 intervals with $1/4$ overlap in (b). The 1-dimensional nerve of $\Ucal$ is the mapper graph in (c). Elements derived from  the mapper graphs are shown in (d).}
    \label{fig:mapper}
    \vspace{-2mm}
\end{figure}

For simplicity, we illustrate the construction of a mapper graph from a 2D point cloud $\Xspace$ sampled from a toy dataset in the shape of a letter `\textit{a}'. 
As shown in~\cref{fig:mapper}(a), $\Xspace$ is equipped with a height function $f$.
To obtain a topological summary of $\Xspace$, we start with a finite cover $\Vcal = \{V_1, V_2, \dots\}$ of $f(\Xspace) \subset \Rspace$ formed by a set of partially-overlapping intervals, such that $f(\Xspace) \subseteq \bigcup_j{V_{j}}$, as shown in~\cref{fig:mapper}(b). 
In other words, the function value $f(x)$ of each point  $x \in \Xspace$ lies within some interval $V_j$. We then consider the clusters induced by points in $f^{-1}(V_j)$ (for each $j$) that are enclosed by rectangles. 
These clusters are denoted as \( \Ucal = \{U_1, U_2, \dots\} \). For instance, the points in \( f^{-1}(V_1) \) form a single cluster \( U_1 \), while those in \( f^{-1}(V_2) \) give rise to two clusters, \( U_2 \) and \( U_3 \), as illustrated in \cref{fig:mapper}(a).

We then construct the 1D \emph{nerve} of $\Ucal$ as the \emph{mapper graph}: each cluster $U_i$ is represented as a node $i$, and there is an edge between node $i$ and node $j$ if $U_i$ and $U_j$ have nonempty intersection.  

The function $f$ is referred to as the \emph{lens}, as it provides a perspective through which we examine the data. Different lens choices, such as density and eccentricity, offer distinct insights \cite{BiasottiGiorgiSpagnuolo2008,SinghMemoliCarlsson2007}. In this work, we adopt the $L_2$-norm as the lens function. The $L_2$-norm of a word embedding captures how strongly an LLM is activated by the input token, and has been shown to yield meaningful results when analyzing hidden representations from deep learning~\cite{RathoreChalapathiPalande2021,RathoreZhouSrikumar2023}. 

A mapper graph is constructed with additional parameters in mind. The cover $\Vcal$ is controlled by the number of intervals $n$ and the overlap $p$. While these values are typically hand-tuned by users, studies have explored automatic parameter selection~\cite{CarriereMichelOudot2018,ChalapathiZhouWang2021}. We use DBSCAN~\cite{EsterKriegelSander1996} as the clustering method, which requires two parameters: minPts (minimum points in a neighborhood to form a core point) and $\varepsilon$ (maximum radius/distance for points to be considered neighbors). {\tool} estimates $\varepsilon$  automatically using the elbow method~\cite{EsterKriegelSander1996,ZhouChalapathiRathore2021} and allows users to set minPts (defaulting to $3$).

In this work, we apply mapper graphs to explore the LLM embedding spaces. We introduce terminology below used by {\tool} to support its exploratory workspace; see~\cref{fig:mapper}(d) for an illustration. 
We use the word ``point'' and ``embedding'' interchangeably. 
We introduce mapper nodes, edges, paths, components, and trajectories, all of which are types of \emph{mapper elements}. 

In a mapper graph, a \mycolorbox{lightblue}{node} represents a set of word embeddings that belong to the same topological neighborhood. 
An \mycolorbox{pastelpurple}{edge} captures the connectivity between two  neighborhoods with nonempty intersection. 
A \mycolorbox{lightorange}{path} between a source node and a target node is the shortest path that connects them in the mapper graph. 
A  \mycolorbox{lightarmygreen}{component} in a mapper graph is a maximal set of nodes in which each pair of nodes is reachable from one another via edges.

As described in \cref{sec:mapper-agents}, exploring a node helps summarizing and annotating a local neighborhood in the embedding space, whereas exploring an edge helps differentiating adjacent neighborhoods. Exploring a path helps capture evolution in the embedding space. For instance, Nicolau et al.~\cite{NicolauLevineCarlsson2011} showed that exploring different ``arms'' in a mapper graph formed by high-throughput microarray data----a form of paths---help understand tumor progression.  

Furthermore, we define a 1-token perturbation of an embedding as the embedding generated by altering a single token in the sentence while preserving the original token; the resulting embedding is referred to as a \emph{perturbed embedding}. 
A \mycolorbox{lightpastelyellow}{trajectory} between an embedding in a source node and a embedding in the target node is formed by a sequence of 1-token perturbations in the embedding space, and visualized by its projection onto the nearest nodes in the mapper graph. Formally, let $x_0 \in S$ be an embedding in a source node, and $x_k$ be an embedding in a target node, $\{x_0, x_1, \dots, x_k\}$ is the sequence of 1-token perturbations forming a trajectory connecting the source with the target. Each perturbed embedding $x_i$ ($0 < i < k$) along the trajectory is projected onto its nearest node in the mapper graph. In short, a trajectory represents a random walk within the embedding space, and visualizing its projection onto the mapper graph reveals an exploratory route among topological neighborhoods of the space. 

An alternative to the classical mapper construction is the \emph{ball mapper}~\cite{Dlotko2019}. It constructs a covering of $\Xspace$ with balls of radius $\varepsilon$ (the same radius parameter in DBSCAN), thus getting rid of the dependency on the lens $f$. The 1D nerve of such a cover is the \emph{ball mapper graph}. Intuitively, a node in the ball mapper represents a \emph{geometric neighborhood} based on the metric structure on $\Xspace$ (as opposed to the topological neighborhood defined by the classical mapper). The same mapper taxonomy applies. In this paper, {\tool} enables the exploration of LLM embedding spaces through both classical and ball mapper graphs, with a primary focus on the former due to its favorable theoretical properties and well-established applications in embedding space analysis (see~\cref{sec:workspace}).

\section{Related Work}
\label{sec:related-work}

In the following, we review related work on embedding visualization, text perturbation and summarization, as well as mapper-based visualization of high-dimensional data. 

\subsection{Embedding Visualization}

Ever since the introduction of static word embedding models, such as Word2Vec~\cite{mikolov2013distributed} and Glove~\cite{PenningtonSocherManning2014}, various visualization techniques have been created to help examine embedding properties. The early tools primarily focused on enabling the exploration of analogies~\cite{liu2017visual} and investigating the local neighborhoods of words~\cite{heimerl2018interactive}.
With the emergence of contextual language models, new aspects of word embeddings, such as contextual variability, became relevant. Consequently, visualization methods evolved to incorporate these aspects, allowing for a more detailed exploration of word meanings in different contexts.
Currently, many approaches focus on exploring contextual word neighborhoods by visualizing them in a scatter plot after applying a dimensionality reduction technique~\cite{boggust2022embedding,sivaraman2022emblaze,sevastjanova2022adapters,sevastjanova2025layerflows,sevastjanova2021explaining}.
A broad overview of these methods is given in the recent STAR paper by Huang et al.~\cite{huang2023VA+embeddings-star}.
The different approaches aim to understand the embedding changes between models~\cite{boggust2022embedding} and layers~\cite{sevastjanova2022adapters,sevastjanova2022lmfingerprints,sevastjanova2025layerflows}, explore the encoded linguistic properties and word semantic meanings~\cite{sevastjanova2022lmfingerprints,berger2020visually,sevastjanova2025layerflows}, and different forms of bias~\cite{sevastjanova2022adapters}.
Others have introduced new visualization designs that use local metrics on embeddings to provide a global overview of similarities and differences between embedding vectors~\cite{heimerl2020embcomp}.

\subsection{Text Perturbation and Summarization}

Text perturbation methods are often used to improve the robustness and interpretability of a model in NLP. These methods can be applied at various levels of linguistic granularity, from character-level to sentence-level transformations, with different perturbation methods \cite{clouatre-etal-2022-local, moradi2021evaluating, romero2024resilient}.

In terms of perturbation generation methods, early approaches often relied on rule-based transformations \cite{moradi2021evaluating}, while others have moved toward automated generation \cite{jin2020bert, garg2020bae}. These methods avoid the limitations of manual rules by relying on the LLM's understanding of context. More recent methods utilize LLMs' generation capabilities. For instance, Qiang et al. \cite{qiang-etal-2024-prompt} used an LLM to generate paraphrased sentences that were used to improve performance of a different model. Varbench \cite{qian2024varbench} addressed data contamination problem of LLMs by dynamically perturbing variables within test cases. 

Perturbation methods are now widely used to assess the stability of explanations and interpretability mechanisms \cite{hsieh2021evaluations}. Marjanovic et al.~\cite{MarjanovicAugensteinLioma2024} evaluated the output and explanation of different LLMs after a perturbation is made to the input. They showed that the robustness per model and per explanation method differs for each perturbation. Instead of random perturbations, SyntaxShap~\cite{amara-etal-2024-syntaxshap} perturbs text by removing words while preserving grammatical coherence. Additionally, perturbations are used in evaluating explanations by comparing attribution scores before and after, where a robust explanation should show minimal score changes. 

In our study, we are interested not only in perturbations, but also in summaries created by LLMs, as these are used in tasks that support experts to extract information from the embedding vectors. Prior research has shown that LLMs can create summaries that align with those written by humans. Large-scale benchmarks, such as those introduced by Zhang et al.\cite{zhang-etal-2024-benchmarking} and Basyal and Sanghvi\cite{basyal2023text}, have demonstrated that instruction-tuned LLMs outperform traditional abstractive summarization models, particularly in zero-shot and few-shot settings. 

Summarizations can integrate human-like structured reasoning into the LLM generation process, as done by Wang et al.~\cite{wang-etal-2023-element}. They propose an element-aware summarization method where LLMs generate summaries using a structured Chain-of-Thought approach. While the summary generation itself is automated, human experts play a role in designing the evaluation framework and assessment of the alignment between generated summaries and reference texts. This structured approach bridges the gap between fully automated summarization and human-guided summary refinement. 
A recent relevant work for our paper is the work by Mozannar et al.~\cite{mozannar2023effective}, where LLMs are utilized to describe image or sentence clusters by contrasting in-cluster and out-of-cluster examples in a human-AI collaborative setting. While our work could also be considered as a human-AI collaboration for text summarization, we propose a different approach via the interplay between the mapper workspace and mapper agents.    

\subsection{Mapper-Based Visualization of High-Dimensional Data}

In this paper, we use the mapper graph to explore the embedding spaces of LLMs. 
Mapper graph is a 1D skeleton of the mapper construction~\cite{SinghMemoliCarlsson2007}, which has been widely applied in data science, including cancer research~\cite{NicolauLevineCarlsson2011,MathewsNadeemLevine2019}, sports analytics~\cite{Alagappan2012}, gene expression analysis~\cite{JeitzinerCarriereRougemont2019}, micro-epidemiology~\cite{Knudson2020}, genomic profiling~\cite{Cho2019}, and neuroscience~\cite{GeniesseSpornsPetri2019,SaggarSpornsGonzalez-Castillo2018}, among others; a comprehensive overview can be found in~\cite{PataniaVaccarinoPetri2017}.  
Mapper graph has also been applied for graph skeletonization~\cite{RosenHajijWang2023}. In the field of deep learning, mapper graphs have been used to capture the topology of hidden representations obtained from image classifiers and LLMs~\cite{RathoreChalapathiPalande2021,RathoreZhouSrikumar2023}, and to study their evolution across layers~\cite{PurvineBrownJefferson2023} and during fine-turning~\cite{RathoreZhouSrikumar2023}. They have also been used to capture characteristics of adversarial attacks~\cite{ZhouZhouDing2023}.

\section{Task Analysis}
\label{sec:task-analysis}

Boggust et al.~\cite{boggust2022embedding} demonstrated that expert users, including data analysts, model developers, and computational linguists, explore embeddings for diverse purposes. Common tasks include assessing the model’s strengths and weaknesses, and evaluating the information it captures across different layers.
Their work indicated that model-driven users---such as model developers---analyze and compare embedding spaces to gain insights into model performance and behavior. In contrast, data-driven users---like computational linguists---examine embedded representations to reveal characteristics of the underlying data. 
However, both groups of users aim to \textbf{generate and verify hypotheses} about specific subjects. 

In this section, we present key investigation tasks of expert users such as linguists, NLP experts, and model developers. 
These tasks are essential for enabling expert users to explore and interpret embeddings effectively, providing insights into structural patterns, relationships, and underlying data distributions and should be facilitated by a workspace that utilizes a mapper graph for embedding analysis. 

\subsection{User Tasks}
\label{sec:user-tasks}

Interpreting the mapper graph is essential to gain insights into the properties encoded in the LLM embedding spaces. 
Its unique structure---nodes as clusters and edges as overlaps---distinguishes its analysis tasks from those of general graphs~\cite{lee2006task} and introduces  interpretive implications.
In the following, we outline five key user tasks and their implications for interpreting mapper graphs.

\noindent\textbf{T1. Investigating \mycolorbox{lightblue}{Node} Characteristics}. As described in \cref{sec:taxonomy}, a mapper node represents a cluster of contextual word embeddings. 
A node can be analyzed to identify common semantic or syntactic features of the clustered tokens and their surrounding contexts. 
This approach is commonly employed in related work, where embeddings are analyzed in relation to the shared characteristics of their surrounding neighborhoods~\cite{sevastjanova2022lmfingerprints,boggust2022embedding}.
Furthermore, analyzing the differences of two nodes can reveal how the LLM distinguishes different linguistic properties.

\noindent\textbf{T2. Investigating \mycolorbox{pastelpurple}{Edge} Characteristics}. As shown in \cref{sec:taxonomy}, an edge in a mapper graph connects two neighboring nodes and represents overlapping data points between these two nodes. 
Analyzing the similarities and differences between neighboring nodes allows the users to understand how the encoded linguistic properties transition from one node to another.
In the case of word embeddings, these differences may reflect subtle linguistic variations, such as the use of specific function words and their placement within a sentence (e.g., the token ``\textit{for}'' used at the beginning vs. in the middle of a sentence).

\noindent \textbf{T3. Investigating \mycolorbox{lightorange}{Path} Characteristics}. We can extend the exploration of the edges to a path connecting multiple nodes to understand how the encoded linguistic properties evolve over a larger data sample. 

\noindent \textbf{T4. Investigating \mycolorbox{lightarmygreen}{Component} Characteristics}. Disconnected components may correspond to distinct semantic topics within the LLM’s embedding space that are more abstract than the node-level characteristics. They could reveal a higher-level information such as shared reference tokens across embeddings within a component.

\noindent \textbf{T5. Investigating \mycolorbox{lightpastelyellow}{Trajectory} Characteristics}. Understanding potential connections between any two nodes—even if they are not directly connected in the mapper graph—can reveal latent transitions or semantic pathways in the embedding space. 
Exploring such trajectories can help uncover how concepts evolve through intermediate regions.

\subsection{Workspace Requirements}\label{sec:workspace-requirements}

In \cref{sec:user-tasks}, we introduce user tasks that are informed by the underlying structure of a mapper graph. 
The prior work has shown that word embeddings encode various linguistic properties, such as part-of-speech, syntactic function, and semantic similarity. Moreover, multiple properties are encoded in the embeddings simultaneously~\cite{rogers-etal-2020-primer,sevastjanova2021explaining,sevastjanova2022lmfingerprints}.
Thus, the explorable space for gaining insights about the LLM learning capabilities is large.
In addition to offering tailored methods for embedding explanation, we must also help users generate insights in a way that is both effective and efficient. 
Thus, a visual analytics workspace that supports the previously listed user tasks has multiple requirements. 

\noindent \textbf{R1. Support Rapid Identification of Regions of Interest.} 
A visual analytics workspace for embedding analysis should provide an overview of high-level mapper properties. 
This can be achieved through automated data annotations that can help users navigate in the space and detect regions for more detailed inspection.

\noindent\textbf{R2. Support Interactive Element Selection and Inspection.} 
To get insights into properties of mapper elements, users should be able to select each element within the workspace and access detailed information about its specific characteristics.

\noindent\textbf{R3. Support Evaluation of Explanation Robustness.} 
To gain insights that reflect the embedding space's true characteristics, the workspace should support users in assessing the robustness of the gained insights. 

\noindent\textbf{R4. Support User-driven Annotations.}
To facilitate interaction provenance and trace the development of insights, users should be able to create annotations linked to the mapper elements they explore.

\section{Mapper Agents}
\label{sec:mapper-agents}

Interpreting a mapper graph that abstracts an embedding space can be both time-consuming and mentally demanding. 
Therefore, computational assistance is essential for uncovering the underlying embedding properties.
We first identify three main \textbf{operations} that can be used to support the identified user tasks from \cref{sec:user-tasks}, i.e., the \emph{summarization}, \emph{comparison}, and \emph{perturbation} operations.
To enable executing them on the mapper's elements, we introduce two types of mapper agents: an \textbf{Explanation Agent} and a \textbf{Verification Agent}.

\subsection{Operations}
We define three operations an agent can perform, which are \emph{summarization}, \emph{comparison}, and \emph{perturbations}. They are explained as follows. 

\para{Summarization.}~The goal of this operation is to generate an understanding of a mapper element, such as the contextual embeddings of a node, an edge, a path, or a component. For example, for a node \emph{summarization} operation, this entails the \emph{summarization} of common characteristics and linguistic properties of the node's embeddings. 

\para{Comparison.}~The goal of this operation is to understand the similarities and differences between two mapper elements. The input for this operation differs from the \emph{summarization} operation (i.e., a bi-variate vs. a uni-variate input): the \emph{comparison} operation requires two elements of the same type as input, whereas the \emph{summarization} operation requires only one. The output of a comparison encompasses the similarities and differences between two elements. For example, a \emph{comparison} applied to a pair of nodes can provide the similarities and differences between contextual embeddings in two different neighborhoods. 

\para{Perturbation.}~The input for this operation can vary, as it can be one or two mapper elements. The aim is to examine and confirm information obtained from previous operations or observations in the mapper graph. For example, with a  node, 1-token \emph{perturbations} could be applied to its enclosing embeddings to access the robustness of the node explanation. The position of the perturbed embeddings can be examined in relation with earlier operations, such as \emph{summarization} or \emph{comparison}. In case of \emph{summarization}, the similarities with and without the inclusion of the perturbed embeddings can be examined. 
With two elements as input, a \emph{perturbation} operation can support in the exploration between the two. As an example, the trajectory between a source and a target node can be explored through 1-token perturbations from one sentence in the source node to another sentence in the target node (see~\cref{sec:taxonomy}).

\subsection{Agent Types}
\begin{figure}[b!]
    \centering
    \vspace{-5pt}
    \includegraphics[width=0.9\linewidth]{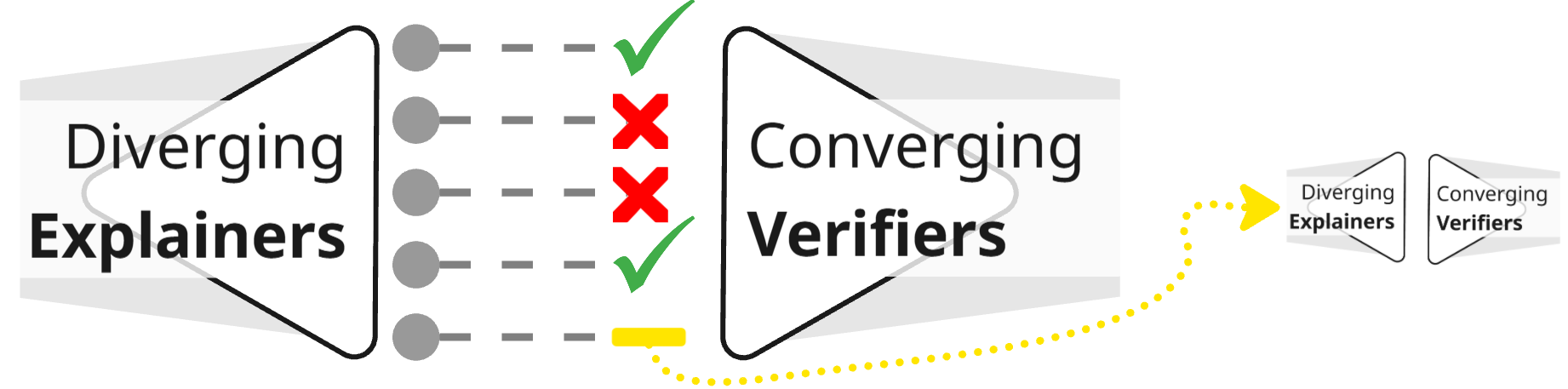}
    \vspace{-5pt}
    \caption{We use the analogy of a decoder-encoder model, where the main purpose of explainers is a divergence (creation of hypotheses) and of verifiers is a convergence (confirmation of hypotheses).}
    \label{fig:decoder-encoder}
\end{figure}

Our main goal is to facilitate the generation and verification of hypotheses about word embedding properties, i.e., to support users to discover possible reasons for the resulting mapper structure (i.e., the unique properties of embeddings within nodes and the characteristics of the transitions between them). 
Thus, we first aim to \emph{diverge} the exploration space, i.e., to create many possible explanation candidates.
In the consecutive step, we aim to \emph{converge} these explanations to those that are robust and, thus, more likely to faithfully explain the underlying embedding space. 
Our main rationale is described in \cref{fig:decoder-encoder} using an encoder-decoder model implemented as two types of agents, i.e., an \emph{Explanation Agent} for hypotheses generation and a \emph{Verification Agent} for explanation (and hypotheses) verification.

\subsubsection{Explanation Agents}
The Explanation Agent class is designed as a set of dedicated \emph{explainers} that utilize the \emph{summarization}, \emph{comparison}, and \emph{perturbation} operations to explain the properties of nodes, edges, paths, components, and trajectories (T1–T5) of a mapper graph.
Our goal is to facilitate hypothesis generation, enabling users to explore diverse insights by generating a wide range of explanations—specifically, identifying potential embedding properties that influence the resulting mapper element structures.
The Explanation Agent class can be implemented in various ways, e.g., utilizing precomputed features and rules as used in previous work \cite{sevastjanova2025layerflows}.
In this paper, we explore the potential of leveraging LLMs to design explainers implemented as tailored prompts for each specific task.
We provide full prompts in the supplement.

\para{\mycolorbox{lightblue}{Node} Explainer.}
The Node Explainer operates on individual nodes to support user task T1 (\cref{sec:user-tasks}). The following table displays the operations together with the input and output that the agent requires to perform the operations. The \emph{summarization} operation analyzes the common linguistic properties of a node's data points, providing insight into the neighborhood of embeddings. The second operation is \emph{comparison}, which shares the input with the \emph{summarization} operation. 
The \emph{comparison} operation presents insights in the embedding neighborhood with a focus on the differences between the data points.  
\captionsetup{labelformat=empty}
\renewcommand{\arraystretch}{1.2}
\begin{table}[h!]
    \vspace{-5pt}
    \centering
    \arrayrulecolor{darkblue} 
    \fcolorbox{darkblue}{lightblue!20}{
    \small{
    \begin{tabular}{|l|p{0.28\linewidth}|p{0.38\linewidth}|}
        \hline
        \rowcolor{lightblue!50} \textbf{Name} & \raisebox{-0.1cm}{\includegraphics[width=0.4cm,height=0.4cm]{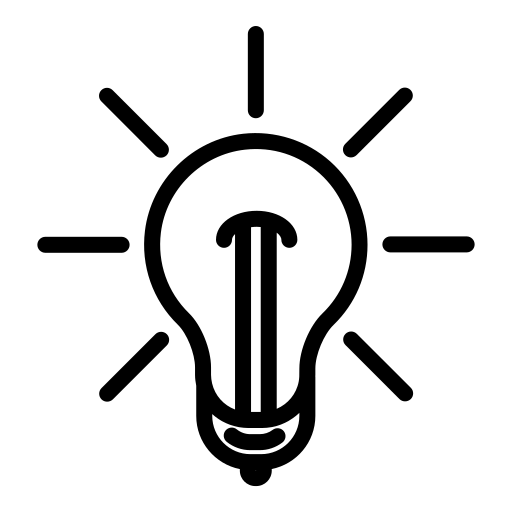}} Node Explainer & \raisebox{-0.1cm}{\includegraphics[width=0.4cm,height=0.4cm]{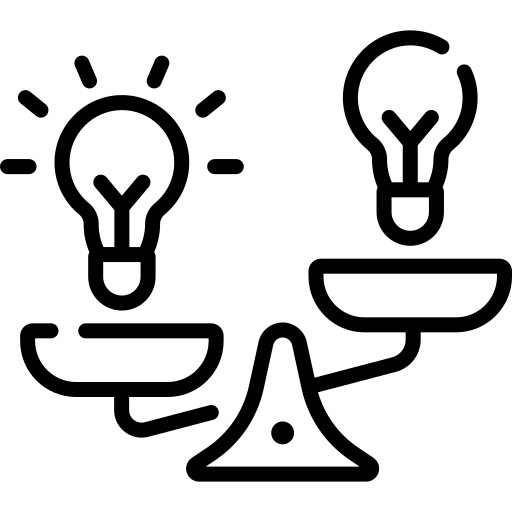}} Node Explainer   \\
        \hline
        \textbf{Operation} & Summarization & Comparison\\
        \hline
        \textbf{Task} & T1 & T1\\
        \hline
        \textbf{Input} & A node and its data points & Two non-adjacent nodes and their data points\\
        \hline
        \textbf{Output} & Common properties of the input data points & Similar and dissimilar properties of the two input node data points\\
        \hline
    \end{tabular}}
    }
    \caption{}
    \vspace{-25pt}
    \end{table}

\para{\mycolorbox{pastelpurple}{Edge} Explainer.}
An Edge Explainer has two nodes as input, which are adjacent and connected by an edge. This edge denotes the word embeddings that are shared between the two input nodes. The Edge Explainer supports user task T2 in~\cref{sec:user-tasks}. 
The \emph{summarization} operation analyzes shared and unique word instances between the two nodes to identify patterns of similarity and divergence, highlighting how one node (neighborhood) shifts into another. In contrast, the \emph{comparison} operation examines differences between two edges to reveal variations in their transition patterns.

\renewcommand{\arraystretch}{1.2}
\vspace{-5pt}
\begin{table}[h!]
    \centering
    \arrayrulecolor{darkpurple} 
    \fcolorbox{darkpurple}{pastelpurple!20}{
    \small{
    \begin{tabular}{|l|p{0.31\linewidth}|p{0.36\linewidth}|}
        \hline
        \rowcolor{pastelpurple!50} \textbf{Name} & \raisebox{-0.1cm}{\includegraphics[width=0.4cm,height=0.4cm]{summarize.png}} Edge Explainer & \raisebox{-0.1cm}{\includegraphics[width=0.4cm,height=0.4cm]{compare.png}} Edge Explainer  \\
        \hline
        \textbf{Operation} & Summarization & Comparison\\
        \hline
        \textbf{Task} & T2 & T2 \\
        \hline
        \textbf{Input} & Unique and common data points of two adjacent nodes &  A pair of unique and common data points of two adjacent nodes \\
        \hline
        \textbf{Output} & Linguistic patterns that illustrate how one node shifts into its adjacent node  & Similarities and differences of properties of the two edge transitions\\
        \hline
    \end{tabular}}
    }
        \caption{}
    \vspace{-25pt}
\end{table}

\para{\mycolorbox{lightorange}{Path} Explainer.}
Unlike the Edge Explainer, the Path Explainer operates on two nodes that do not have to share overlapping embeddings.
It takes the shortest path between two non-adjacent nodes as input and analyze the data points within each node along the path. 
The \emph{summarization} operations examine how word usage patterns shift, develop, or stabilize as they propagate through the path.
The \emph{comparison} operations compare the evolution patterns between two different paths. 
The output in both cases supports the user in task T3.

\vspace{-5pt}
\renewcommand{\arraystretch}{1.2} 
\begin{table}[h!]
    \centering
    \arrayrulecolor{darkorange}
    \fcolorbox{darkorange}{lightorange!20}{
    \small{
    \begin{tabular}{|l|p{0.32\linewidth}|p{0.35\linewidth}|}
        \hline
        \rowcolor{lightorange!50} \textbf{Name} & \raisebox{-0.1cm}{\includegraphics[width=0.4cm,height=0.4cm]{summarize.png}} Path Explainer  & \raisebox{-0.1cm}{\includegraphics[width=0.4cm,height=0.4cm]{compare.png}} Path Explainer  \\
        \hline
        \textbf{Operation} & Summarization & Comparison \\
        \hline
        \textbf{Task} & T3 & T3\\
        \hline
        \textbf{Input} & Data points along the shortest path between two non-adjacent nodes &  Data points of a pair of path. \\
        \hline
        \textbf{Output} & A summary of progression of properties over the path & Similarities and differences of properties of two paths  \\
        \hline
    \end{tabular}}
    }
        \caption{}
    \vspace{-25pt}
\end{table}

\para{\mycolorbox{lightarmygreen}{Component} Explainer.} 
The Component Explainer operates on connected components of a mapper graph, supporting user task T4 as described in~\cref{sec:user-tasks}. 
The \emph{summarization} operations analyze the points within a component to identify their common linguistic characteristics, uncovering higher-level patterns beyond the individual node level.
The \emph{comparison} operations reveal the similarities and differences between two components. 
 
\renewcommand{\arraystretch}{1.2} 
\begin{table}[h!]
    \centering
    \arrayrulecolor{darkarmygreen} 
    \fcolorbox{darkarmygreen}{lightarmygreen!20}{
    \small{
    \begin{tabular}{|l|p{0.31\linewidth}|p{0.35\linewidth}|}
        \hline
        \rowcolor{lightarmygreen!50} \textbf{Name} & \raisebox{-0.1cm}{\includegraphics[width=0.4cm,height=0.4cm]{summarize.png}} Component Explainer & \raisebox{-0.1cm}{\includegraphics[width=0.4cm,height=0.4cm]{compare.png}} Component Explainer   \\
        \hline
        \textbf{Operation} & Summarization & Comparison \\
        \hline
        \textbf{Task} & T4 & T4 \\
        \hline
        \textbf{Input} & One component & Two components \\
        \hline
        \textbf{Output} & Common properties of nodes within the component & Similarities and differences between nodes of two components   \\
        \hline
    \end{tabular}}
    }
    \caption{}
    \vspace{-25pt}
\end{table}

\para{\mycolorbox{lightpastelyellow}{Trajectory} Explainer.}
The Trajectory Explainer uses the \emph{perturbation} operation before the \emph{summarization}. It displays the trajectory that results when applying a sequence of minor (e.g., 1-token) perturbations to convert one sentence belonging to the source node to a sentence belonging to the target node. The trajectory is created by connecting the perturbed embeddings that are attached to their nearest mapper nodes (i.e., the most similar cluster of embeddings, see \cref{sec:taxonomy}).

\vspace{-5pt}
\renewcommand{\arraystretch}{1.2}
\begin{table}[h!]
    \centering
    \arrayrulecolor{darkpastelyellow} 
    \fcolorbox{darkpastelyellow}{lightpastelyellow!20}{
    \small{
    \begin{tabular}{|l|p{0.7\linewidth}|}
        \hline
        \rowcolor{lightpastelyellow!50} \textbf{Name} & \raisebox{-0.1cm}{\includegraphics[width=0.4cm,height=0.4cm]{summarize.png}} Trajectory Explainer  \\
        \hline
        \textbf{Operation} & Perturbation + Summarization \\
        \hline
        \textbf{Task} & T5 \\
        \hline
        \textbf{Input} & Two data points, one from each of the two selected nodes   \\
        \hline
        \textbf{Output} &  A trajectory that connects the two data points and the perturbed intermediate sentences attached to their most similar node/edge   \\
        \hline
    \end{tabular}}
    }
        \caption{}
    \vspace{-30pt}
\end{table}

\subsubsection{Verification Agents}
After creating a set of explanation candidates for the different mapper elements, we utilize Verification Agents to inspect which of the explanations are robust and thus more likely to faithfully explain the embedding properties. 
In other words, our goal is to converge the explanation space to the most reliable candidates.
Since an Explanation Agent is implemented using LLM prompts, this verification step addresses an important challenge of LLMs being prone to hallucinations.
Similarly to the Explanation Agent class, this class utilizes the three operations, i.e., \emph{summarization}, \emph{comparison}, and \emph{perturbation} and is implemented using an LLM and tailored prompts, provided in the supplement.

\para{\mycolorbox{lightblue}{Node}, \mycolorbox{pastelpurple}{Edge}, \mycolorbox{lightorange}{Path}, \mycolorbox{lightarmygreen}{Component} Verifier.}
A Verification Agent uses the \emph{perturbation} operation to verify earlier obtained outputs from the Explanation Agents. 
The following table shows how a Verification Agent can be applicable to verify the explanations of the Node Explainer, but similar procedure is used for the Edge, Path, and Component Verifier. 
The input for the Verification Agent is determined by the scope of the Explainer to which the verification is applied (e.g., a node and its summary for the Node Verifier, and an edge and its summary for the Edge Verifier). 
We then utilize the \emph{perturbation} operation to generate perturbed sentences for the input data points, which rephrase the original sentence or introduce a 1-token change while preserving the target token.
By default, five perturbed sentences are created for each input data point utilizing an LLM.
The particular prompt used is provided in the supplement.
We then select one perturbed sentence for each original sentence in a node that still remains within the same node, by computing the average distance between its word embedding and all word embeddings in the node. If this distance is smaller than the average pairwise distance within the node, we retain this perturbed sentence.   
A new explanation is created on the perturbed examples using the same method as for the original Explainer.
We then evaluate the similarity between the original and the new explanation applying a cosine similarity on the sentence embedding vectors extracted using the MiniLM model~\cite{wang2020minilm}. 
The similarity is used as an indicator of the explanation's robustness.

\captionsetup{labelformat=empty}
\renewcommand{\arraystretch}{1.2} 
\begin{table}[!ht]
    \centering
    \arrayrulecolor{darkblue} 
    \fcolorbox{darkblue}{lightblue!20}{
    \small{
    \begin{tabular}{|l|p{0.33\linewidth}|p{0.35\linewidth}|}
        \hline
        \rowcolor{lightblue!50} \textbf{Name} & \raisebox{-0.1cm}{\includegraphics[width=0.4cm,height=0.4cm]{summarize.png}} Node Verifier &\raisebox{-0.1cm}{\includegraphics[width=0.4cm,height=0.4cm]{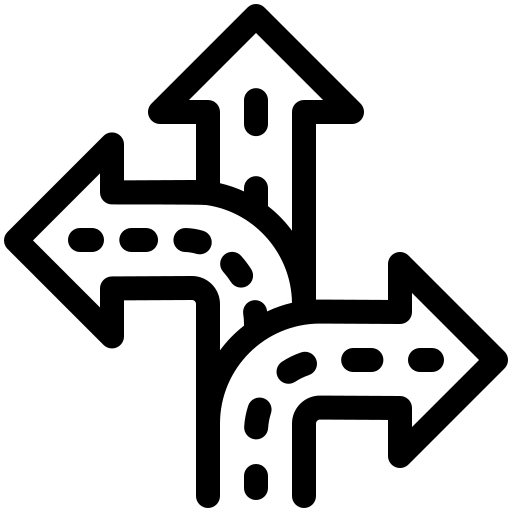}} Node Verifier  \\
        \hline
        \textbf{Operation} & Perturbation + Summarization & Perturbation + Comparison\\
        \hline
        \textbf{Task} & T1 & T1\\
        \hline
        \textbf{Input} & Perturbed data points of a target node, the initial summary & Perturbed data points of two target nodes, the initial comparison\\
        \hline
        \textbf{Output} & Summary of the perturbed elements and its similarity to the initial summary & Comparison of the perturbed elements and its comparison to the initial comparison\\
        \hline
    \end{tabular}}
    }
    \caption{}
        \vspace{-30pt}
\end{table}

\captionsetup{labelformat=default}
\captionsetup{labelformat=default}
\begin{figure*}[t!]
    \centering
    \includegraphics[width=\textwidth]{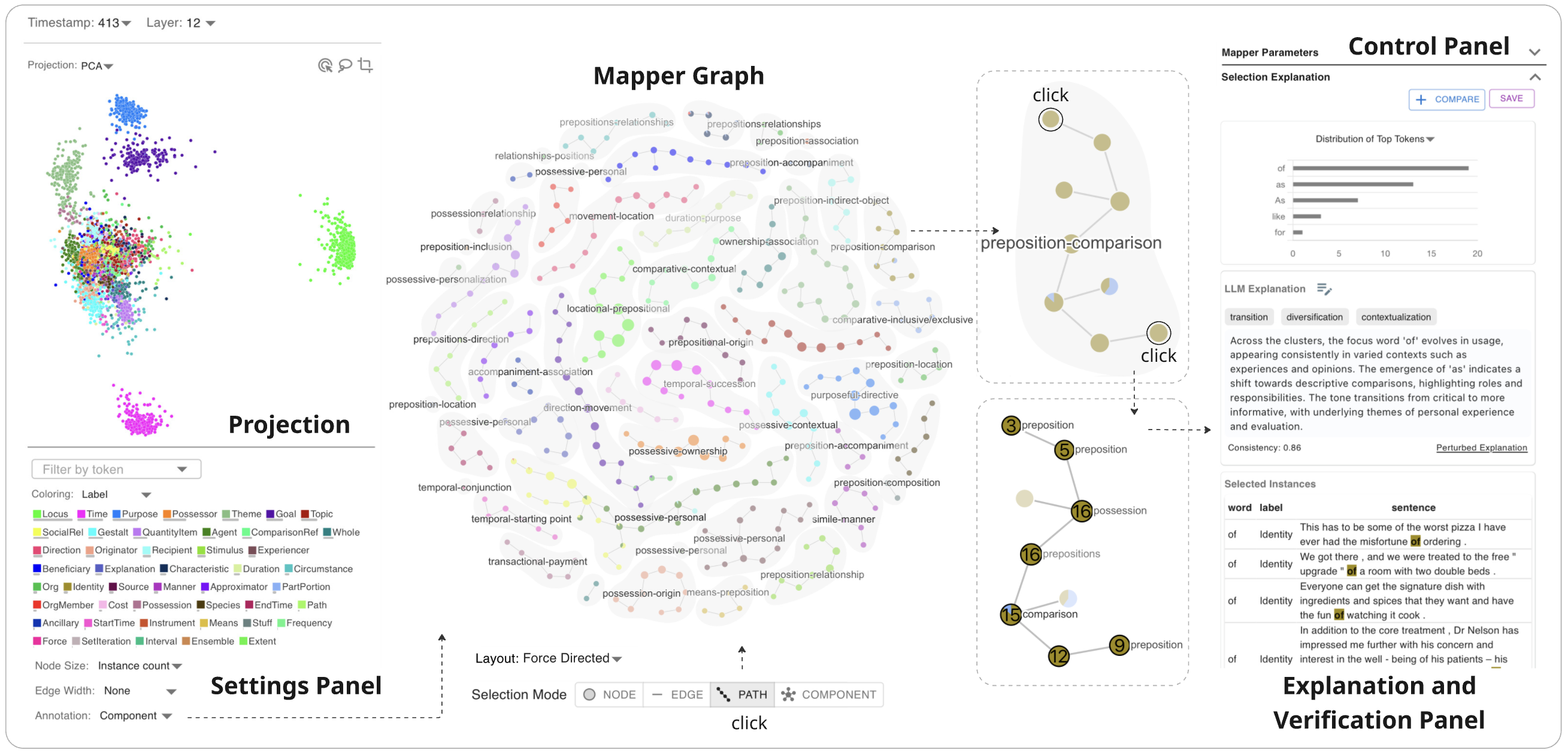}
     \vspace{-17pt}
    \caption{The {\tool} workspace consists of two main visualizations, i.e., the \emph{Mapper Graph} and \emph{Projection} visualized as a scatter plot. Properties associated with the mapper graph and its underlying embeddings are displayed in the \emph{Settings Panel}. The user can specify mapper parameters in the \emph{Control Panel}. To explore the visualized embedding properties, the users can select the mapper elements to explain, where explanations are followed by verifications, in the \emph{Explanation and Verification Panel}.}
    \label{fig:workspace}
    \vspace{-15pt}
\end{figure*}

\begin{figure}[h!]
\vspace{-5pt}
    \centering
    \includegraphics[width=0.9\linewidth]{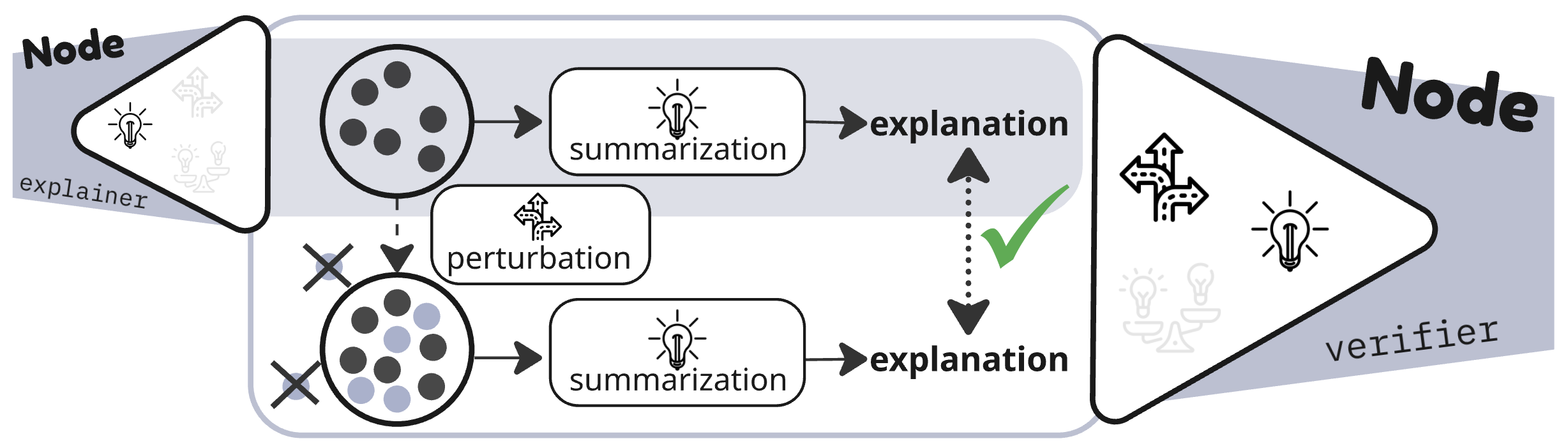}
    \vspace{-5pt}
    \caption{A Node Verifier takes a node (with its data points) and the initial explanation created by a Node Explainer as input. It creates perturbations of the input data points, and creates a new explanation of the perturbations that fall inside the input node's neighborhood. The Verifier measures the similarity between the initial and the new explanations.}
    \vspace{-5pt}
    \label{fig:verifier-process}
\end{figure}

For instance, consider a Node Verifier that employs the \emph{summarization} operation. This agent receives a node along with its summary as input. We introduce perturbations (i.e., 1-token substitutions and rephrasings) to the original sentences of the node, identify which of the perturbed sentences fall within the same neighborhood as the original ones, and generate a new summary based on these perturbed sentences, as showcased in~\cref{fig:verifier-process}. The similarity between the original and perturbed summaries reflects the robustness of the explanation.

\captionsetup{labelformat=default}

\para{\mycolorbox{lightpastelyellow}{Trajectory} Verifier.}
The verification of the Trajectory Explainer is an exceptional case. 
Due to its unique implementation which utilizes perturbations to connect two nodes in the space, it is challenging to verify its robustness automatically, thus, the verification of this explainer is done by the user.

\section{Mapper Workspace}
\label{sec:workspace}

\begin{figure}[b!]
    \centering
        \vspace{-20pt}
    \includegraphics[width=\linewidth]{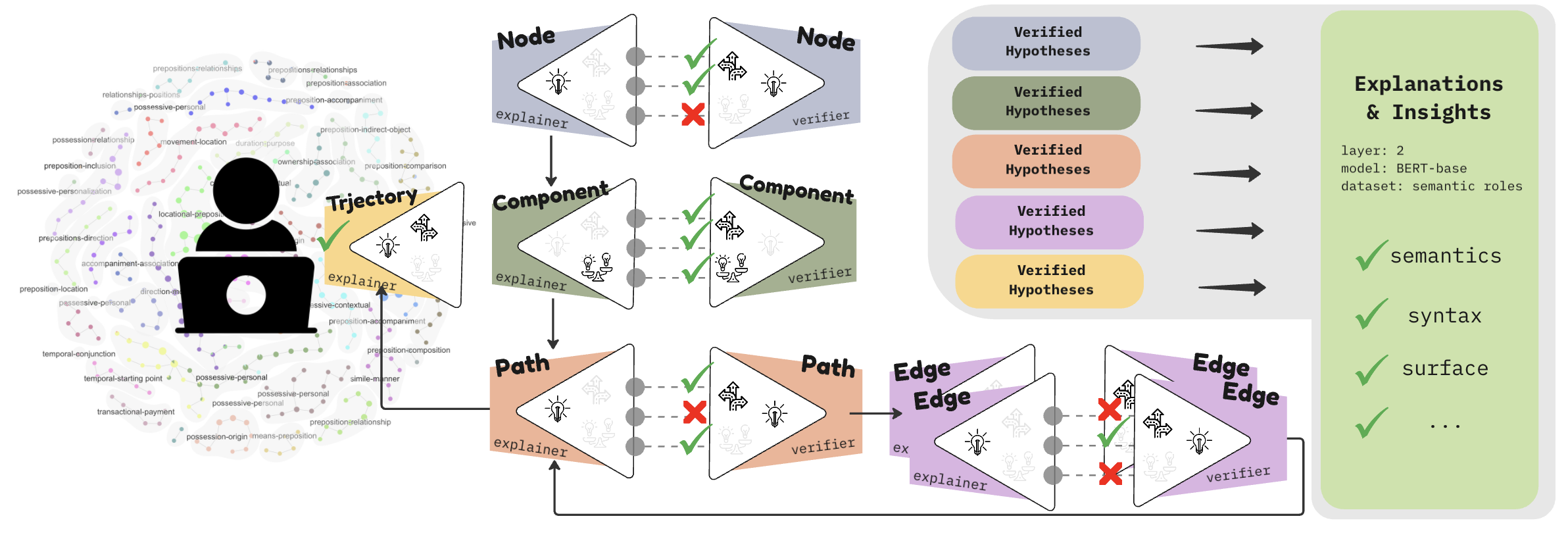}
    \vspace{-15pt}
    \caption{We support an iterative process involving numerous combinations of Explainers and Verifiers to generate insights regarding linguistic properties of contextual word embeddings.}
    \label{fig:workflow}
\end{figure}

To assist users in generating and verifying hypotheses about embedding properties using the agents introduced in \cref{sec:mapper-agents}, we present the {\tool} workspace, as shown in \cref{fig:workspace}.
The workspace supports the tasks introduced in \cref{sec:user-tasks} and addresses the identified requirements in \cref{sec:workspace-requirements}.
In particular, the proposed workspace enables an iterative embedding exploration by combining different agents to generate insights, as exemplified in \cref{fig:workflow}. 
When starting with the analysis, the users can gain an overview of the explorable space through precomputed mapper component annotations (\textbf{R1}), select mapper elements for closer investigation (\textbf{R2}), create their explanations (\textbf{T1-T5}), and verify the explanation robustness (\textbf{R3}). 
The workspace supports note-taking and interaction provenance (\textbf{R4}).

\para{Data Preprocessing.}
The data preprocessing steps involve extracting token embeddings layer-by-layer from an LLM on a sufficiently large text corpus (e.g., 12 layers of a BERT-base model) and associating each token with various linguistic features, such as its semantic role and part-of-speech (POS) tag. In addition, each data point in the mapper graph is equipped  with its $L_2$-norm. 

\begin{figure*}[t!]
    \centering
    \includegraphics[width=\textwidth]{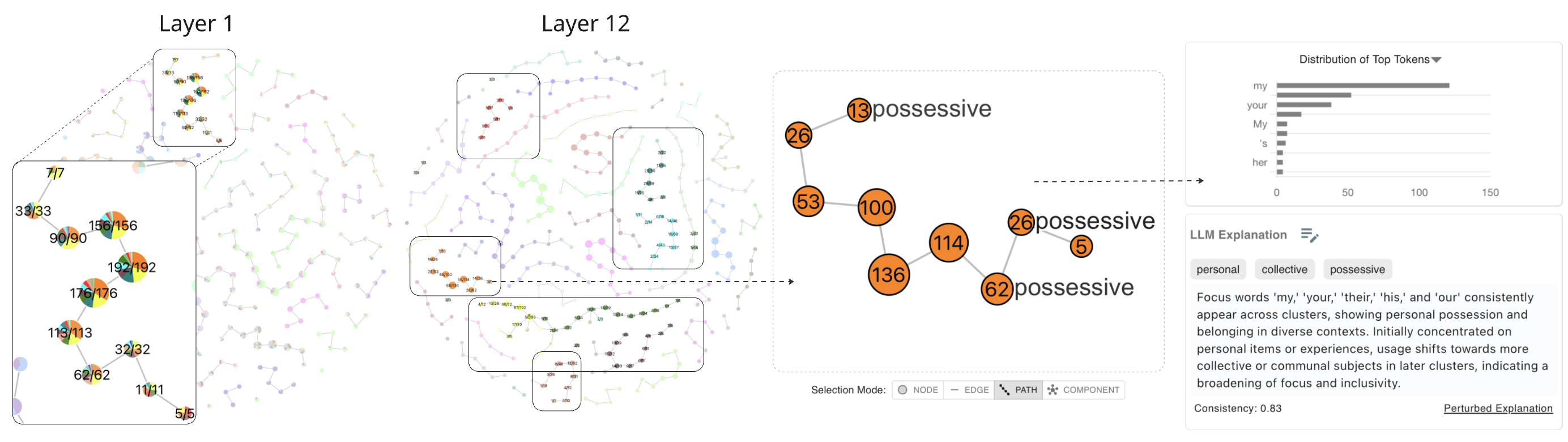}
    \vspace{-20pt}
    \caption{Exploring occurrences of the token  `\textit{my}.' In layer 1, embeddings remain noncontextualized and all tokens are clustered into a single component. In layer 12, the embeddings are contextualized to the specific classification task for which the model has been fine-tuned. The node annotations emphasize the token common characteristic, i.e., that they are \textit{possessive} pronouns. The Path Explainer provides more insight into the property transition along the path, i.e., the tokens are used in contexts ranging from ``\textit{personal experience}'' to a ``\textit{collective subject}.''}
    \label{fig:my}
    \vspace{-12pt}
\end{figure*}

\subsection{Control Panel and Settings Panel}

In the \textit{control panel}, the users specify the parameters for the mapper algorithm, introduced in \cref{sec:taxonomy}.
They can create the \textit{classical mapper} or the \textit{ball mapper}. For the classical one, the users specify the parameters for the DBSCAN clustering algorithm (i.e., the \textit{minPts} and $\varepsilon$). Other parameters for the mapper are \textit{cover number} and \textit{cover overlap}. 
For the ball mapper, only the $\varepsilon$ parameter needs to be specified by the user. We compute the mapper graph as described in \cref{sec:taxonomy} on embeddings extracted from a single layer at a time, as specified by the user.

In the \textit{settings panel}, the users can specify the features for designing the mapper graph, e.g., an attribute to be mapped to the color/size of a node, or the width of an edge; see \cref{sec:mapper-graph-panel} for details. 
After computing the graph, we generate explanations for the components and nodes as representative keywords and their verification scores. These keywords and scores are used as default annotations in the interface, specified by users in the settings panel. 
Additionally, the panel allows users to filter specific tokens for more detailed exploration.

\subsection{Mapper Graph}
\label{sec:mapper-graph-panel}

The mapper visualization utilizes the design from the original \textit{Mapper Interactive}~\cite{ZhouChalapathiRathore2021} work that introduced the toolbox for visual exploration of mapper graphs .
The nodes in the graph represent embedding clusters and the edges the overlapping data points between two connected nodes.
The default graph layout is created using a force-directed layout.
The users can change the layout to an \emph{anchored}  version where nodes are positioned based on their centroids in a projected 2D space. 
A node is visualized either as a circle that displays the $L_2$ norm of the cluster, or as a pie chart, where the slices represent one of the associated linguistic properties (e.g., the semantic roles or POS tags). 
The size of a node can scale with the number of data points it encloses, or it can remain constant. 
The width of an edge can encode the Jaccard similarity between adjacent nodes, or it can stay constant.   

\para{Interaction Methods.}
Similar to \textit{Mapper Interactive}, the workspace enables basic interactive operations such as zooming, dragging, and panning.
The users can enable keyword-based annotations by specifying it in the settings panel.
The precomputed explanation keywords are attached to the mapper elements, whereby their opacity represents the verification scores (the higher the score, the higher the opacity).
By default, the keywords are attached to the components. 
To reduce keyword overlap, we prioritize those with higher verification scores while hiding overlapping ones.
When zoomed in, the annotations adjust to the keywords associated with various nodes and more keywords appear.

The selection panel beneath the mapper graph allows users to select different types of mapper elements (i.e., nodes, edges, paths, and components) and their underlying data points through four data selection modes. Users can make a direct selection by interacting with elements in the mapper graph view. Once a selection is made, users can apply an Explainer to generate a summary of an element or compare explanations of two elements of the same type. Additionally, users can generate a trajectory explanation by selecting a sentence from the source node and target node and specifying the number of 1-token perturbations connecting them.

\subsection{Projection}
In addition to the mapper graph, we display embeddings in a scatter plot utilizing a projection method (e.g., PCA, MDS, t-SNE, or UMAP) specified by the user in the interface.
The projection can be used for navigation, i.e., the users can select clusters in the projection to locate them in the mapper graph as well as for designing an anchored layout for the mapper graph.
Data points (i.e., embeddings) are displayed as dots and colored according to the selected (linguistic) property (i.e., semantic role, POS tag, or $L_2$-norm).

\para{Interaction Methods.} 
The users can select a data point by clicking on the corresponding dot in the visualization or select a group of data points using the lasso or brushing functionality. We use a linking and brushing method, i.e., the clusters of the selected data points are highlighted in the mapper graph and vice versa.

\subsection{Explanation and Verification Panel}
When the user selects one or two mapper elements from the selection panel below the mapper graph, a bar chart is displayed showing the token or label frequencies within the selected elements.
Additionally, the tokens and their corresponding sentences are displayed in a table format within the \textit{explanation and verification panel}. Moreover, an explanation is generated, and its verification is automatically performed using the mapper agents introduced in \cref{sec:mapper-agents}.

\para{Explanation Design.} 
A single explanation consists of a natural language description and a list of three descriptive keywords that summarize the selected element(s).
The user can change the Explainer's operation from summarization to comparison by clicking on the \textit{compare} button on top of the respective panel.
For comparison, the user needs to select another element of the same type in the mapper graph to serve as the reference.
Once the second element is selected, the explanation is recomputed and displayed in the particular panel.
Every time an explanation is created (either utilizing the summarization or comparison operator), the explanation keywords are attached to the particular element in the mapper graph.

\para{Verification Design.} An explanation's robustness is highlighted through a consistency score displayed underneath the explanation, whereby it is executed automatically every time a new explanation is created.
Users can explore the reason for the particular score by clicking on the \textit{perturbed explanation} button. 
Then, the panel displays the explanation generated for the perturbed examples, and the table shows these examples for a closer inspection.

\para{Note Taking and Visual Annotations.}
To help users gain an overview of the explained elements and the corresponding insights, the workspace allows for note-taking.
Specifically, when an explanation is generated, users can modify it by adding additional observations or removing irrelevant parts of the auto-generated explanations.

Furthermore, when a Trajectory Explainer is applied, the system projects the trajectory onto nearest nodes or edges of the mapper graph. 
For each perturbed example, we compute its embedding and locate the nearest data point  (within the $\varepsilon$ threshold) in the graph sharing the same $L_2$-norm interval (see~\cref{sec:background}). 
The perturbed example is then assigned to a node or edge based on the location of its nearest neighbor. 
We visualize the trajectory by placing points on the corresponding nodes or edges and connecting them in order, see~\cref{fig:trajectory} for an example. 
Users can also refine the trajectory by inserting or deleting perturbed examples.

\begin{figure}[t!]
    \centering
    \includegraphics[width=\linewidth]{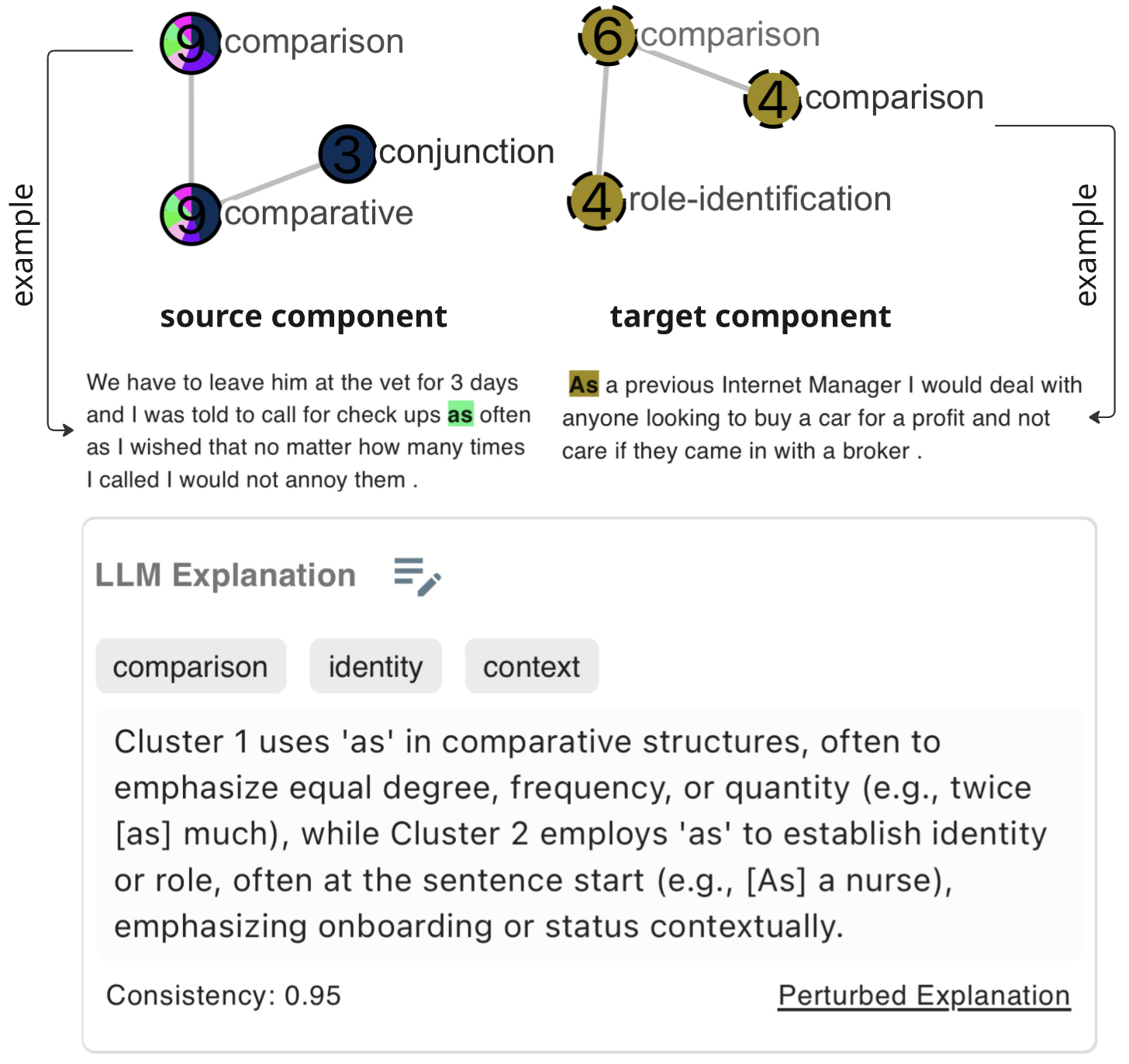}
    \vspace{-15pt}
    \caption{Layer 5 of the fine-tuned BERT captures the two functions of the token `\textit{as}', i.e., its role of conjunction (connecting clauses or phrases) and preposition (by showing a relationship between that noun/pronoun and another word in the sentence). The two functional roles get separated into two distinct connected components.}
    \label{fig:as}
    \vspace{-10pt}
\end{figure}

\section{Case Studies}
\label{sec:case-studies}

In the following, our goal is to evaluate the effectiveness of the proposed methodology. Specifically, the Explanation and Verification Agents integrated into the {\tool} workspace, in supporting the generation of reliable insights into embedding properties. To achieve this, we present replication studies aimed at reproducing insights from previous work regarding the embedding properties encoded in different layers of the BERT-base model. 

\subsection{Models and Data}
The authors of TopoBERT fine-tuned multiple versions of the BERT model on three NLP tasks, generating word embeddings across various layers and fine-tuning steps~\cite{RathoreZhouSrikumar2023}. 
They then applied the mapper algorithm to these embeddings to examine and compare the fine-tuning processes.
Following the approach in TopoBERT, we fine-tune the BERT-base model on a NLP task called \textit{preposition supersense disambiguation}, that classifies prepositions into their semantic categories. Specifically, we use the \textit{Supersense Role} annotations from the STREUSLE v4.2 dataset~\cite{schneider2015corpus}, which provides 41 labels across 4,282 preposition tokens. 
We fine-tune the BERT-base model for 7 epochs and then extract 768-dimensional word embeddings from each of its 12 layers. 
Finally, mapper is applied to the embeddings of each layer. 
GPT-4o~\cite{hurst2024gpt} serves as the underlying model for the agents.

\subsection{Prior Insights on BERT's Embeddings}
Extensive research has focused on understanding the embedding properties of BERT, with the majority of studies examining its pre-trained version.
Rogers et al.~\cite{rogers-etal-2020-primer} showed that BERT captures linguistic information (e.g. parts of speech, syntactic chunks, and semantic roles) at different layers of the model's architecture. 
The early layers of BERT encode surface features such as word lexical similarity~\cite{sevastjanova2022lmfingerprints} as well as sentence length and presence of words ~\cite{jawahar-etal-2019-bert}.
The early to middle layers encode the syntactic properties of the word, in particular, the basic syntactic information, i.e. POS tags ~\cite{tenney-etal-2019-bert, liu-etal-2019-linguistic}. 
Semantic information seems to be encoded throughout the whole model~\cite{tenney-etal-2019-bert}, whereby the upper layers encode task-specific information~\cite{liu-etal-2019-linguistic}.
Moreover, TopoBERT~\cite{RathoreZhouSrikumar2023} has shown embedding patterns in the fine-tuned model through two examples: a shift from fronted to non-fronted token usage along a path, and the grouping of identical tokens despite differing semantics.

Next, we demonstrate that our agent-based approach replicates both well-known patterns of the pre-trained model and previously explored patterns in the TopoBERT.
More cases are offered in the supplement.

\subsection{Replication Study}
\para{Possessive Pronouns.} Using the Node Explainer, we investigate how the fine-tuned BERT learns possessive pronouns. We first filter the pronoun `\textit{my}' and display its occurrence in the mapper graph in layers 1 and 12. 
In layer 1, where embedding vectors remain noncontextualized, all tokens are clustered into a single component (see~\cref{fig:my}).
In layer 12, the embeddings are contextualized to the specific classification task for which the model has been fine-tuned.
In this layer, tokens are distributed across multiple components, grouped together with other related tokens.
By using a different explainer, the Path Explainer, on the two nodes at the opposite ends of the component, we can gain insights into the transition of the encoded properties. 
In particular, the explanation says: \textit{focus words `my,' `your,' `their,' `his,' and `our' consistently appear across clusters, showing personal possession and belonging in diverse contexts. Initially concentrated on personal items or experiences, usage shifts towards more collective or communal subjects in later clusters, indicating a broadening of focus and inclusivity}.
The Path Verifier produces a high consistency score (0.83) that indicates that the explanation of the perturbed examples of the path's underlying data points produces a similar explanation to the initial one.

\begin{wrapfigure}[17]{r}{0.21\textwidth} 
  \centering
  \vspace{-15pt}
  \includegraphics[width=\linewidth]{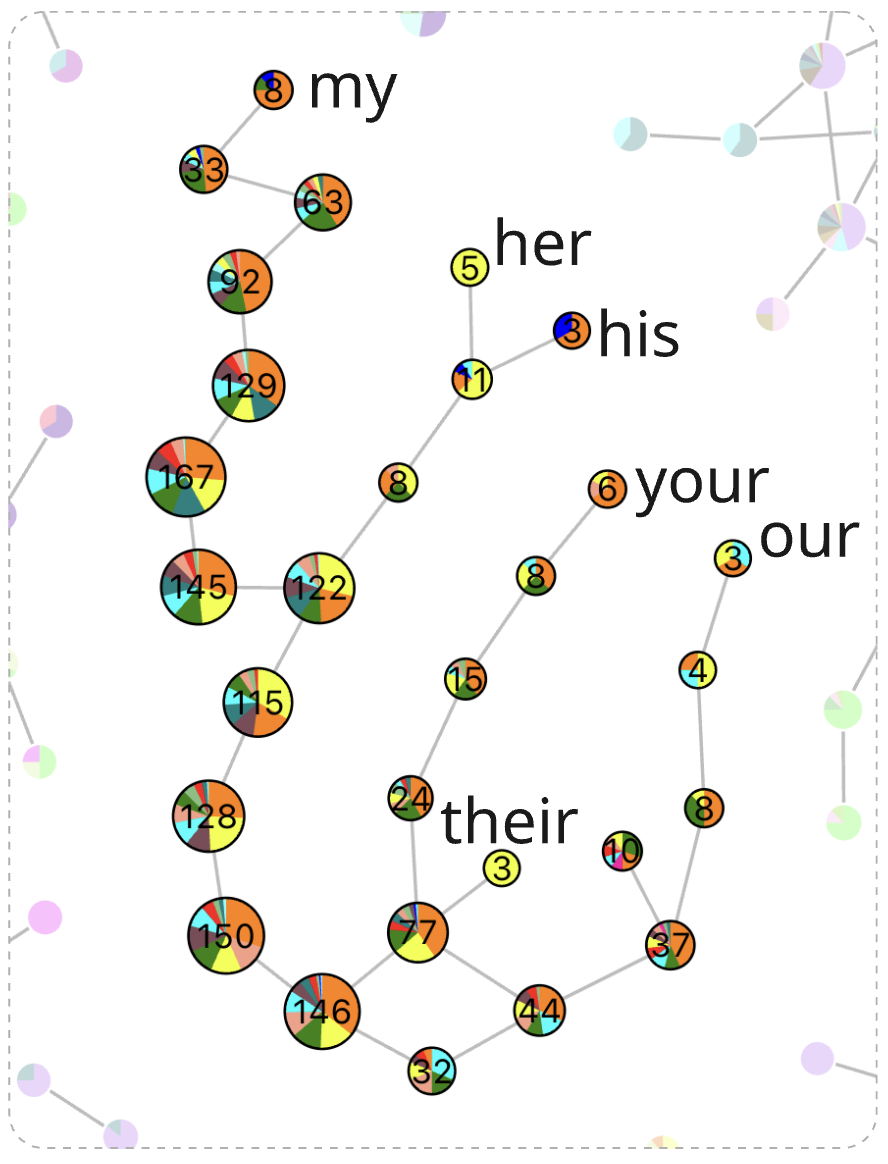}
  \vspace{-15pt}
  \caption{BERT embeddings for pronouns in layer 4.}\label{fig:pronous}
\end{wrapfigure}

It has been shown that the middle layers (3-6) of BERT encode token syntactic functionality while also being semantically contextualized~\cite{rogers-etal-2020-primer}.
To verify this hypothesis, we explore the properties of possessive pronouns (e.g., `\textit{my}', `\textit{her}', `\textit{his}') in layer 4.  
As shown in~\cref{fig:pronous}, all of them are part of a single component. 
By iteratively applying the Node Explainer, we generate explanations for the terminal nodes of the resulting branches within the component.
We observe that each branch corresponds to a unique possessive pronoun, indicating that the hypothesis is valid, i.e., in middle layers, tokens encode their function while being semantically contextualized.

\begin{figure}[t!]
    \centering
    \includegraphics[width=\linewidth]{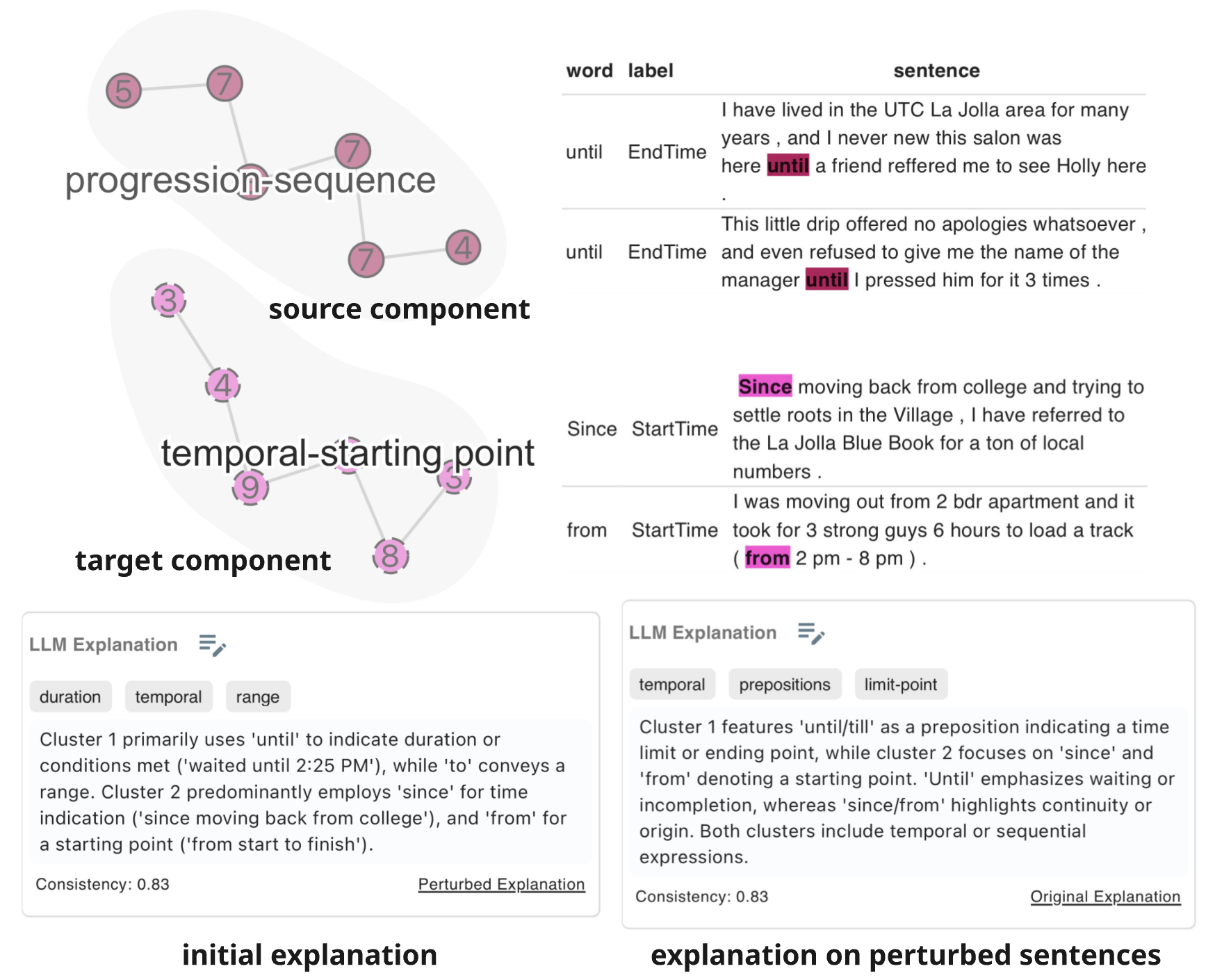}
    \vspace{-17pt}
    \caption{We use the Component Explainer to compare the two components of temporal preposition tokens. The summary, that can be verified by the Component Verifier, states that the two components have in common that they both highlight temporal sequences but differ in focus: the source component focuses on initiation and continuity in time, while the target component on culmination and time limits.}
    \label{fig:time}
    \vspace{-15pt}
\end{figure}

\para{\textit{As} in Conjunction and Preposition.}
The token `\textit{as}' can function as either a conjunction or a preposition, depending on its usage in a sentence. 
As a conjunction, it connects clauses or phrases, showing a relationship between them. 
As a preposition, it introduces a noun or pronoun, showing a relationship between that noun/pronoun and another word in the sentence. 
In~\cref{fig:as}, we show how the fine-tuned BERT captures these syntactic properties in its middle layers.
In particular, in layer 5, the tokens get separated into two distinct components. 
We examine their linguistic properties by annotating component nodes with descriptive keywords. These annotations suggest two distinct embedding patterns: one component primarily represents conjunctions, while the other relates to ``role-identification.'' To explore this distinction further, we use the Component Explainer to compare the two components.
The explanation provides insight into the main differences. One component primarily uses `\textit{as}' in comparative structures to indicate equal degree, frequency, or quantity (e.g., ``\textit{twice [as] much}''), while the other employs `\textit{as}' to denote identity or role, often at the beginning of a sentence (e.g., ``\textit{[As] a nurse}''), emphasizing status or on-boarding. 
This distinction is in accordance with the functional role of prepositions.
The Component Verifier produces a high consistency score, which is shown by the resemblance of the explanations of the perturbed components to the initial explanations. 

\para{Temporal Prepositions and Trajectory Exploration.}
Different temporal prepositions (e.g., `\textit{since}', `\textit{from}', `\textit{until}') exist that can be used to indicate when an action or event takes place.
In layer 12, we observe that these prepositions are contextualized such that tokens indicating the start of an action form a distinct component, separate from those representing the end time.
We use the Component Explainer to compare the two components.
The summary, which can be verified by the Component Verifier with a high consistency score, states that the two components share the common trait of highlighting temporal sequences but differ in focus: the source component focuses on initiation and continuity in time, while the target component focuses on culmination and time limits.
The components, corresponding token sentences, and explanations are provided in~\cref{fig:time}.

\begin{figure}[t!]
    \centering
    \includegraphics[width=\linewidth]{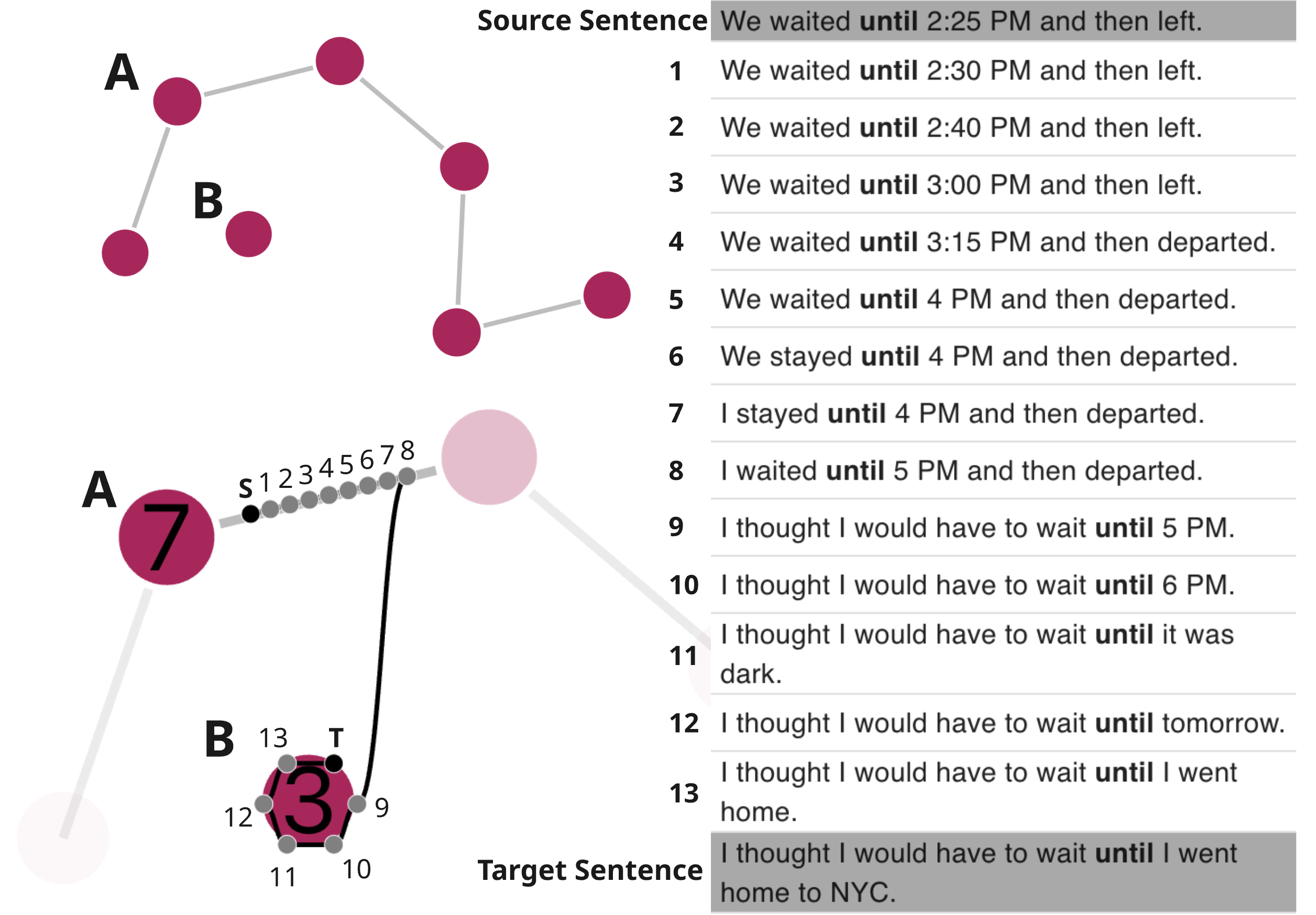}
    \vspace{-1.2em}
    \caption{Trajectory exploration from node A to node B in layer 12 of the fine-tuned BERT. A source and a target sentence are selected from each node, and 13 intermediate sentences are generated by LLM and projected onto the graph.}
    \label{fig:trajectory}
    \vspace{-2em}
\end{figure}

As shown in~\cref{fig:trajectory}, tokens labeled \textit{Endtime} form two disconnected components, with one containing only a single node (B). To investigate this separation, we construct a perturbation trajectory connecting the two components. Specifically, we select nodes A and B as endpoints, as both contain only the focus word `\textit{until}.' From each node, we manually choose a representative sentence as the source and target, respectively. By specifying the number of perturbations, LLM generates 13 intermediate sentences that gradually transform from the source to the target. These sentences are then projected onto the mapper nodes or edges based on their word embeddings and arranged sequentially along the trajectory path.
We observe that the transition between nodes A and B occurs between sentences 8 and 9. Notably, in sentences 1–8, `\textit{until}' appears earlier in the sentence (e.g., ``\textit{We waited until...}''), while in sentences 9–13, it appears toward the end (e.g., ``\textit{...waited until...}''). This suggests that word position may influence the separation in embedding space.
Furthermore, while the LLM suggests that the key difference between nodes A and B lies in the use of ``precise times,'' we observe that sentences 9 and 10 also contain precise times yet remain associated with node B, indicating that this hypothesis does not hold.
\section{Discussion and Future Work}

Our study introduces {\tool}, a framework that aims to assist users in exploring and verifying explanations for elements in mapper graphs using different agents. Through our evaluation, we demonstrated how the Explanation and Verification Agents facilitate the understanding of complex embedding structures by guiding users through a process of divergent explanation generation and convergent verification. With this approach, we encourage the user to move toward an interactive AI-assisted exploration. By enabling users to systematically assess linguistic properties within embeddings, our method offers a structured way to generate and validate hypotheses about word embeddings.

\para{LLM Hallucinations.}
A known problem of LLMs is the potential risk of hallucinations \cite{huang-hallucinations-2025}, where the model generates credible-sounding but incorrect 
answers. Although we try to mitigate this risk with the Verification Agent, its effectiveness is constrained by the inherent limitations of the LLMs. The Verification Agent assesses the robustness~\cite{hedstrom2023quantus} of the explanation through perturbations, providing insight into its stability, though it does not guarantee absolute correctness.

To reduce the impact of hallucinations in the future, an approach could be an explanation ensemble, where multiple LLMs or different prompting strategies are used to cross-check and validate explanations, highlighting inconsistencies that may indicate hallucinations. Confidence scores can be assigned to the explanations, or the variance between the different explanation outputs can be assessed. 

Another direction of future work to mitigate the risk of hallucinations is the incorporation of human feedback. The LLM can be requested to display its  Chain-of-Thought for a generated explanation, where humans can rate or modify the result. This refines the process of the Explanation Agent, asking humans to correct the wrong explanations. 

\para{Explanation Verification.} 
In addition to the limitations of LLMs, there are limitations related to the perturbation methods. 
The perturbation strategy primarily relies on 1-token modifications, assuming that minor changes have minimal impact on embeddings and, consequently, on the explanations. 
However, depending on the sentence and the position or type of perturbation, small changes can have a large impact on the sentence embedding. In contrast, in other cases, large changes only have a small impact on the embedding. This can lead to wrong assumptions about the robustness of the explanation in both directions. Investigation of the perturbation strategy by incorporating multi-token modifications or context-aware substitutions could refine verification and better capture the nuances of embedding behavior.

Additionally, the trajectory is verified by the user, which deviates from the intended scalable and automated approach. Future work could investigate a way to automate the Trajectory Verification process. One such way, which is in line with the current verification strategies, is to generate multiple trajectories between nodes with different perturbations while maintaining semantic coherence. A similarity score can be computed between these trajectories, and if multiple perturbed trajectories lead to similar transitions between nodes, this increases the confidence in the validity of the trajectory. 

Another way that does not include the generation of new trajectories could be the incorporation of graph-based validation strategies. The trajectory is analyzed to ensure that perturbations follow the expected path between two nodes of interest. 

\section{Conclusion}
\label{sec:conclusion}

In this paper, we utilize mapper graphs to explore the topological structures of LLM embedding spaces. To facilitate the analysis of mapper elements, we introduce {\tool} workspace, accompanied by two types of mapper agents, i.e., Explainers and Verifiers. These agents employ summarization, comparison, and perturbation operations to generate and validate explanations of element properties, including linguistic aspects of clusters, transitions, and connectivity. 
We first \textit{diverge} the exploration space, i.e., to create explanation candidates.
In the consecutive step, we \textit{converge} these explanations to those that are robust and, thus, more likely to faithfully explain the underlying data. 
Through multiple case studies, we show how the {\tool} workspace can be used to replicate findings from prior work. 
The workspace will be made publicly available under acceptance.

\section*{Acknowledgment}
This work was partially supported by grants from the U.S.~National Science Foundation (projects IIS-2145499, IIS-2205418, and DMS-2134223) and the Swiss National Science Foundation (project 10003068).

\bibliographystyle{abbrv-doi-hyperref}
\bibliography{refs-explainable-mapper}

\appendix
\section*{Supplement}
In this supplement, we first present additional use cases for the classical mapper (\cref{supp:more-cases}) and several examples using ball mapper (\cref{supp:ball-mapper}). We then provide examples from the Groningen Meaning Bank dataset~\cite{Bos2017} (\cref{supp:GMB}). 
Finally, for reproducibility, we provide a complete list of the prompt templates used in the paper, along with examples of prompt engineering (\cref{supp:prompts}).

\section{Additional Use Cases for the Classical mapper}
\label{supp:more-cases}

Using {\tool}, we present additional use cases for the classical mapper applied to the fine-tuned BERT model, as described in the main paper. 

\begin{figure}[!ht]
    \centering
    \includegraphics[width=\linewidth]{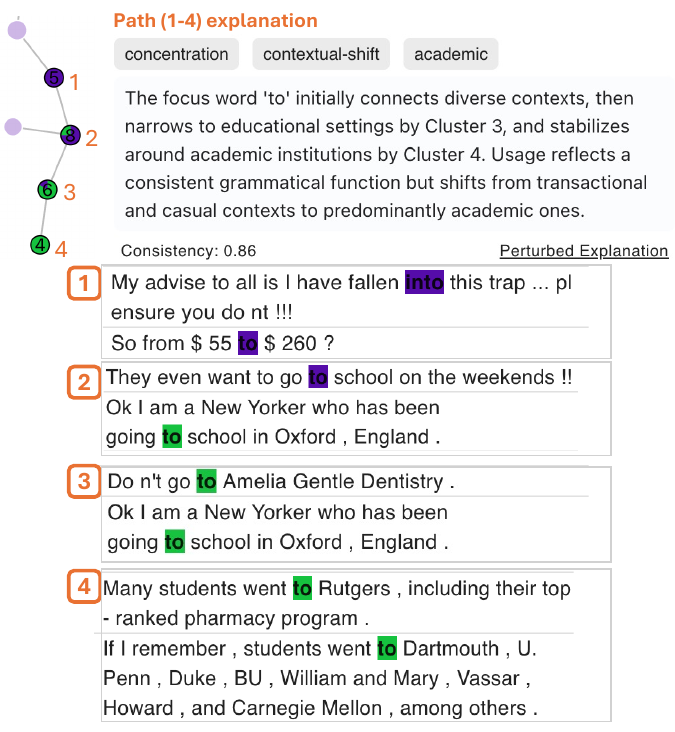}
    \vspace{-6mm}
    \caption{A path in the final layer of the fine-tuned BERT model, where supersense-role labels evolve from “Goal” (purple) to “Organization” (green). The path explanation and example instances for each node are provided.}
    \label{fig:confusion-path}
    \vspace{-2mm}
\end{figure}

\para{Model Confusion about Supersense-Role Label.} 
In the final layer of a BERT model fine-tuned on the supersense-role task, most components in the Mapper graph cluster by supersense-role labels. However, some nodes exhibit a mix of labels, indicating instances where the model is confused or uncertain. For example, along the path shown in~\cref{fig:confusion-path}, node 1 labeled “Goal” gradually transitions to node 4 labeled “Organization.” We utilize the Path Explainer and obtain an explanation with a high consistency score of 0.86. The explanation reveals that the path focuses on the token \textit{`to'}, which initially appears in diverse contexts, then narrows to educational settings, and eventually stabilizes around academic institutions. This my suggest that the model struggles to differentiate \textit{`to'} between labels “Goal” and “Organization” in educational contexts.

\begin{figure}[!ht]
    \centering
    \includegraphics[width=\linewidth]{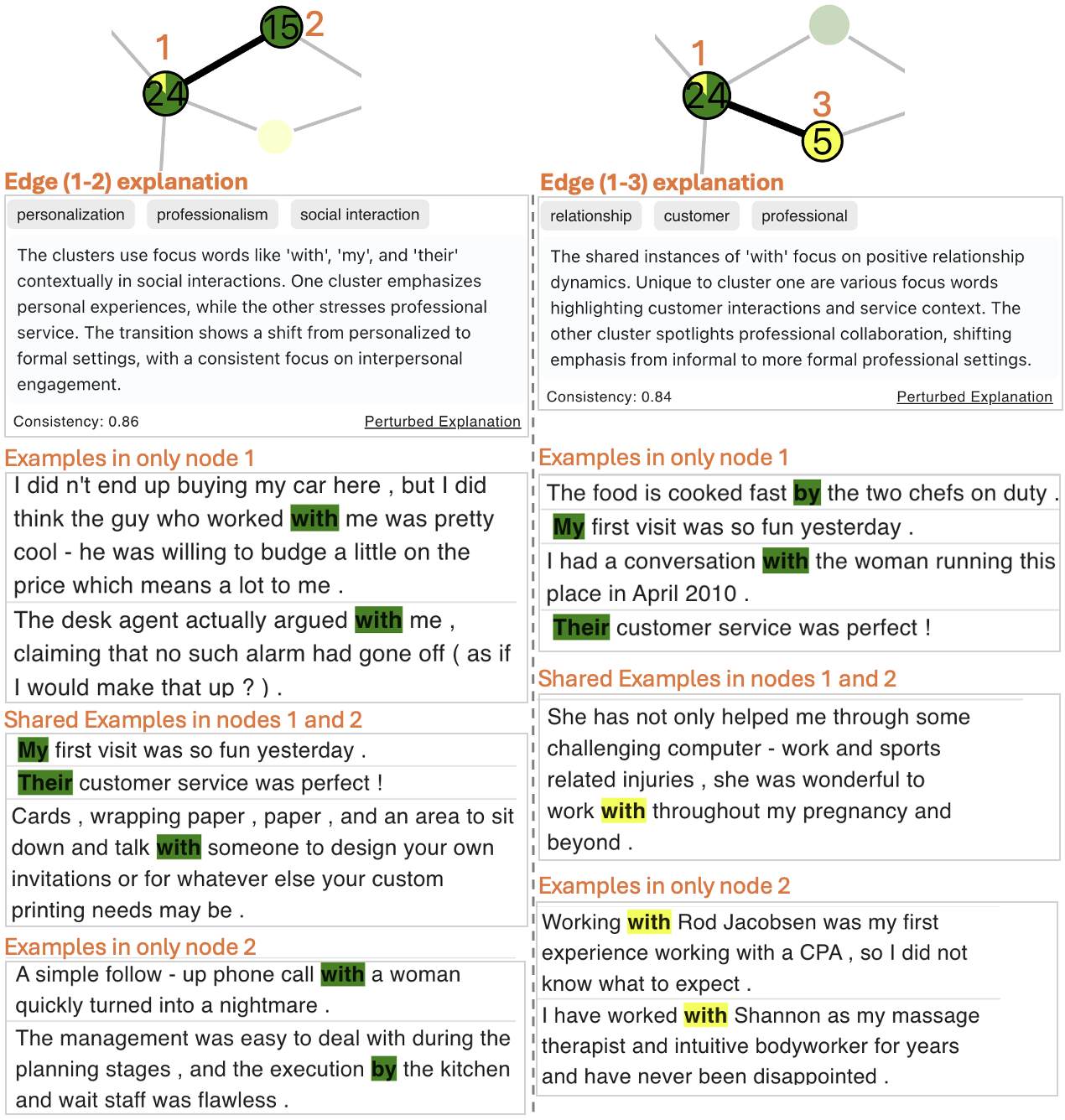}
    \vspace{-6mm}
    \caption{Two edges originating from the same mapper node in the final layer of a fine-tuned BERT model, the edge explanations, and instances in edges.}
    \label{fig:two-edge explanation}
    \vspace{-4mm}
\end{figure}

Another example is shown in~\cref{fig:two-edge explanation}, where a node with mixed labels---``Agent” (green) and ``SocialRel” (yellow)---is separated into two pure nodes. We apply the Edge Explainer to the two edges and obtain two edge explanations with high consistency scores. The edge from node 1 to node 2 reflects a shift from personal experiences to professional services within social interaction contexts. In contrast, the edge from node 1 to node 3 captures a transition from customer interactions and services to professional collaborations (\textit{e.g., “work with…”}). This suggests that the model is learning to differentiate between social and professional contexts in nuanced ways.

\begin{figure}[!ht]
    \centering
    \includegraphics[width=\linewidth]{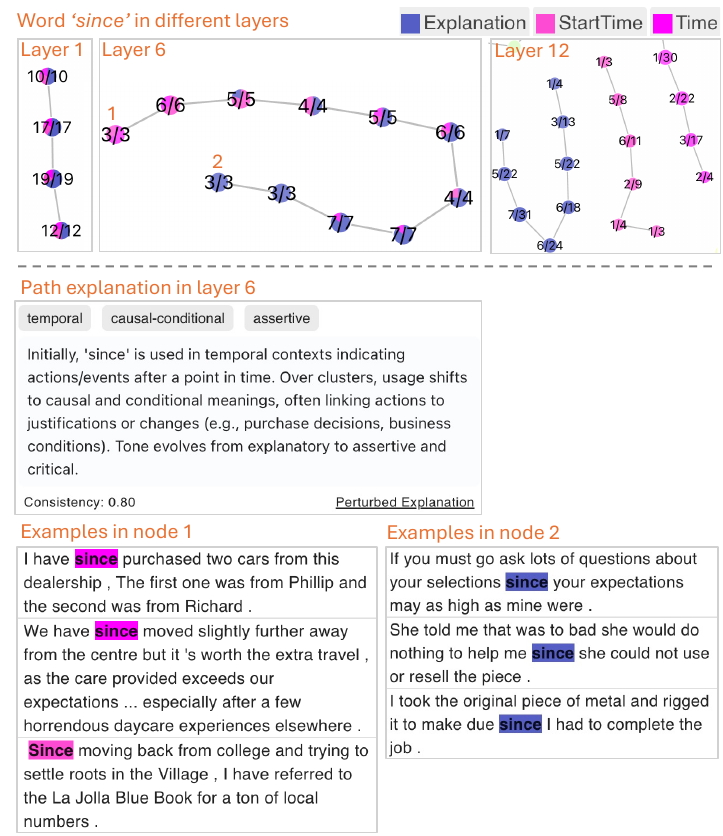}
    \vspace{-4mm}
    \caption{The word \textit{`since'} across layers 1, 6, and 12, along with its path explanation in layer 6.}
    \label{fig:since}
\end{figure}

\para{\textit{Since} in Temporal and Causal Contexts.} 
The word \textit{`since'} can represent temporal information or causality. 
We examine how the usage of \textit{`since'} evolves across layers. 
By searching for the token, as shown in~\cref{fig:since}, we observe that \textit{`since'} instances with different labels are mixed together in layer 1. 
These instances form a path that transitions from temporal labels to the “Explanation” label in layer 6. 
By layer 12, they are separated into three distinct components, aligning with their respective labels.

\begin{figure}[!ht]
    \centering
    \includegraphics[width=\linewidth]{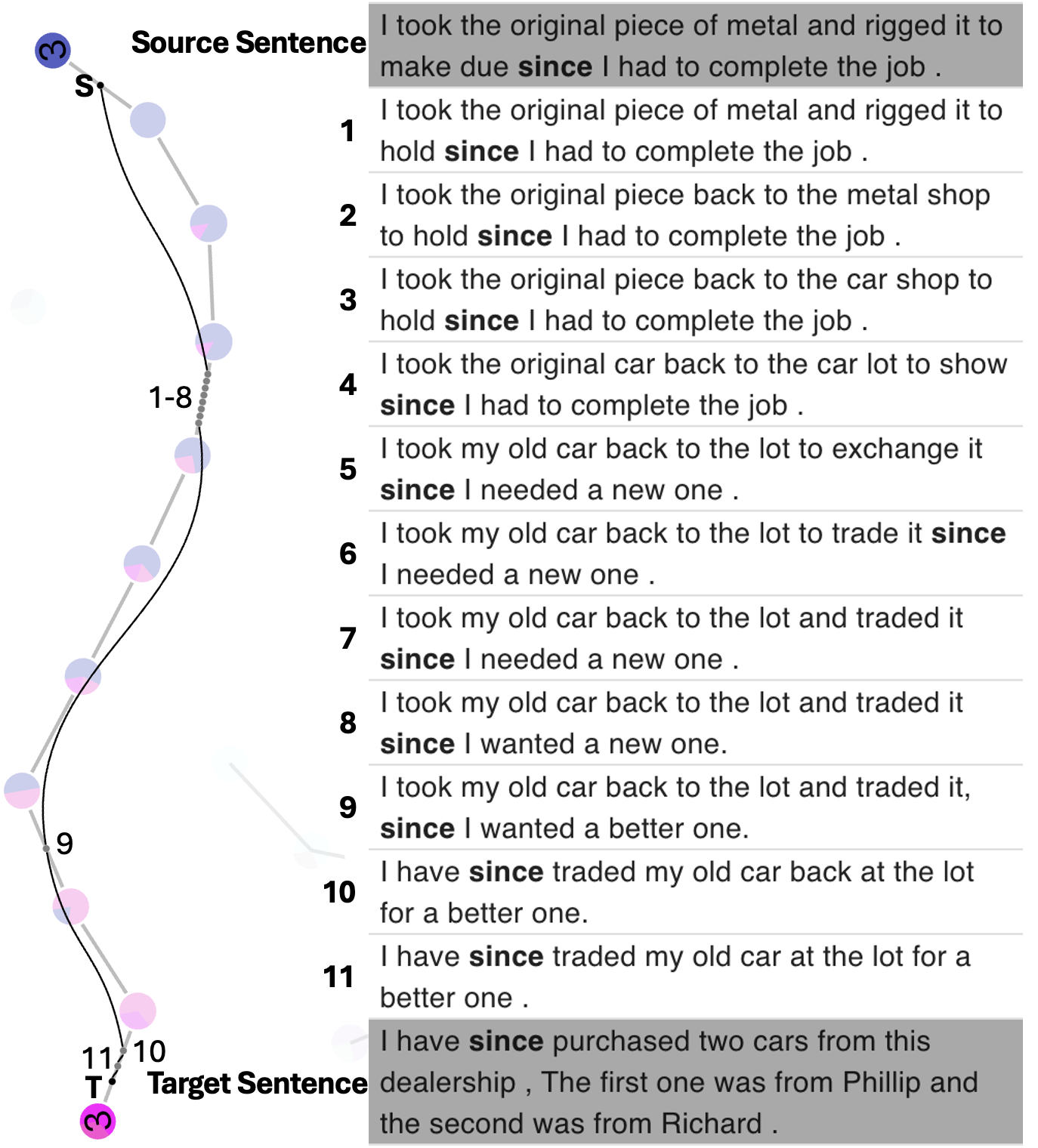}
    \vspace{-4mm}
    \caption{The trajectory between two end nodes of the \textit{`since'} path formed in layer 6.}
    \label{fig:since-trajectory}
\end{figure} 
We then examine the path evolution in layer 6 using the Path Explainer, which reveals that \textit{`since'} is initially used in temporal contexts to indicate actions or events following a point in time. 
As the path progresses, its usage shifts toward causal and conditional meanings. 

To further analyze this transition, we apply the Trajectory Explainer by constructing a trajectory between the two end nodes of the path.
As shown in~\cref{fig:since-trajectory}, the trajectory generally follows the path, where \textit{`since'} expresses a causal meaning in sentences 1–9 and shifts to a temporal information in sentences 10 and 11. The attachment positions of sentences 9 and 10 reflect a shift from a mixed mapper node to a mapper node with pure temporal labels, strengthening the hypothesis of path evolution.

\section{Examples for the ball mapper}  
\label{supp:ball-mapper}
\begin{figure}[!ht]
    \centering
    \includegraphics[width=\linewidth]{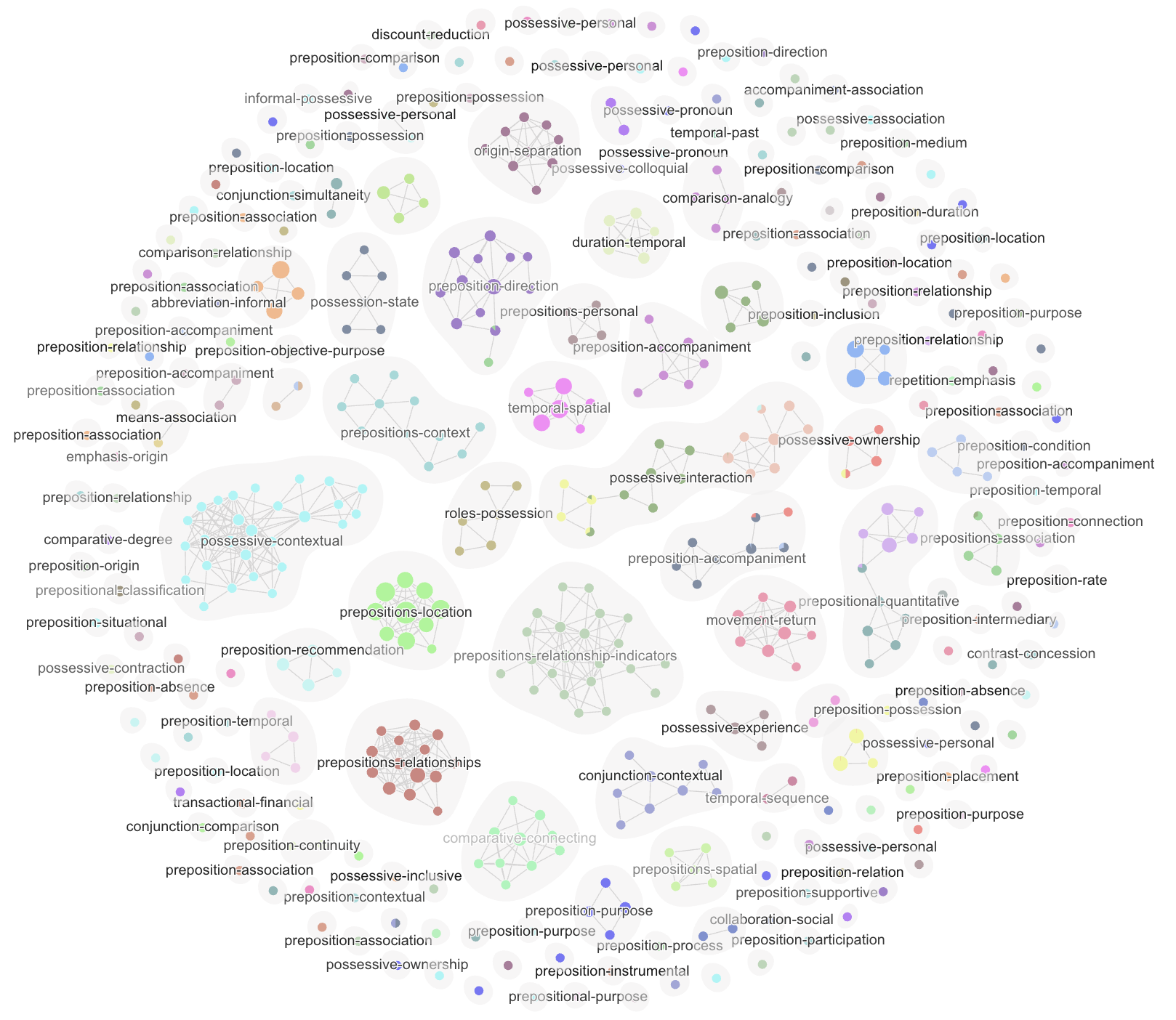}
    \caption{Ball mapper of the final layer of the fine-tuned BERT model, with component annotations overlaid.}
    \label{fig:ball-mapper}
\end{figure}

In this section, we present several examples of {\tool} based on the ball mapper. \cref{fig:ball-mapper} provides an overview of the ball mapper applied to the final layer of a fine-tuned BERT model, where we use the same $\varepsilon$ value as in the classical mapper to define the geometric neighborhood.

\begin{figure}[!ht]
    \centering
    \includegraphics[width=\linewidth]{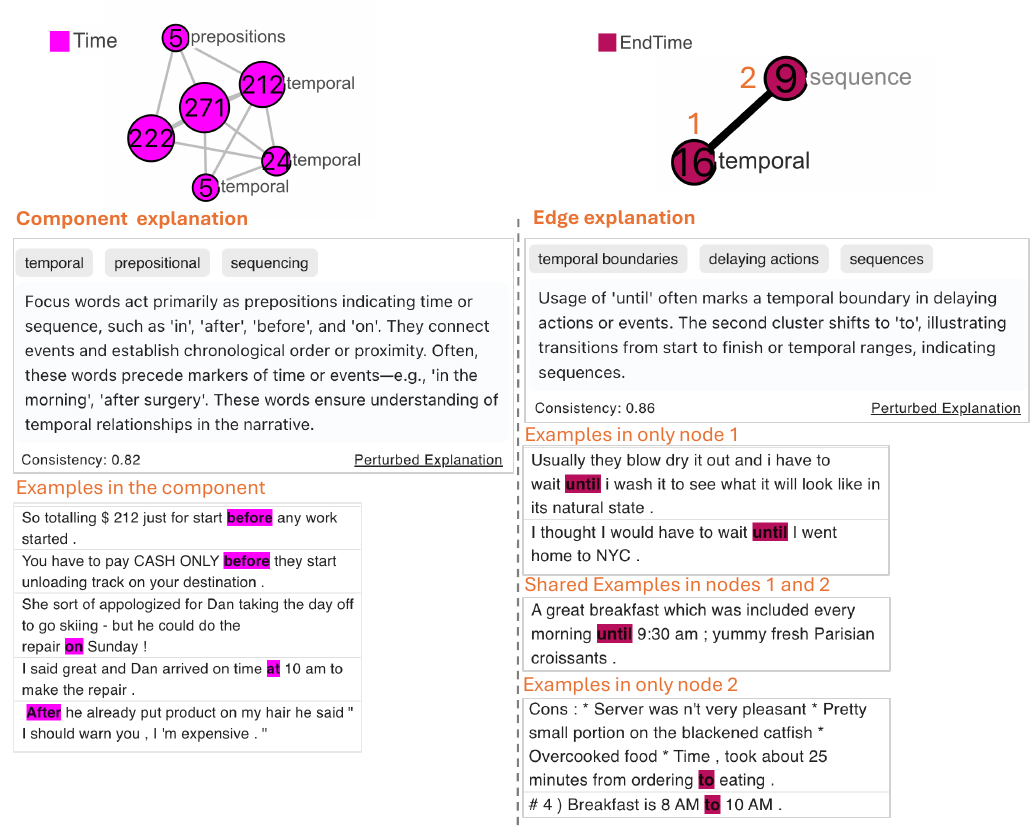}
    \vspace{-6mm}
    \caption{A component (left) and a path (right) in the ball mapper.}
    \label{fig:ball-mapper-2cases}
    \vspace{-4mm}
\end{figure}

\para{Component Investigation.}
In the ball mapper, we observe that tokens are generally grouped into components based on their supersense-role labels, with more complex inner connections compared to the classical mapper.
For example, \cref{fig:ball-mapper-2cases} (left) shows a component composed of tokens labeled ``Time.'' 
The Component Explainer produces a highly consistent explanation, indicating that words such as \textit{`in,'} \textit{`after,`} and \textit{`before'} often precede makers of time events to convey temporal information. 
Similarly, as shown in~\cref{fig:ball-mapper-2cases} (right), the words labeled ``EndTime'' form a single component composed of two nodes. 
To examine its internal structure, we apply the Edge Explainer to characterize the transition between the nodes. 
The explanation reveals a lexical shift from \textit{`until'}, which denotes temporal boundaries (\textit{e.g., ``until I went home''}), to \textit{`to'}, which indicates a time sequence (\textit{e.g., ``8 AM to 10 AM''}).

\begin{figure}[!ht]
    \centering
    \includegraphics[width=\linewidth]{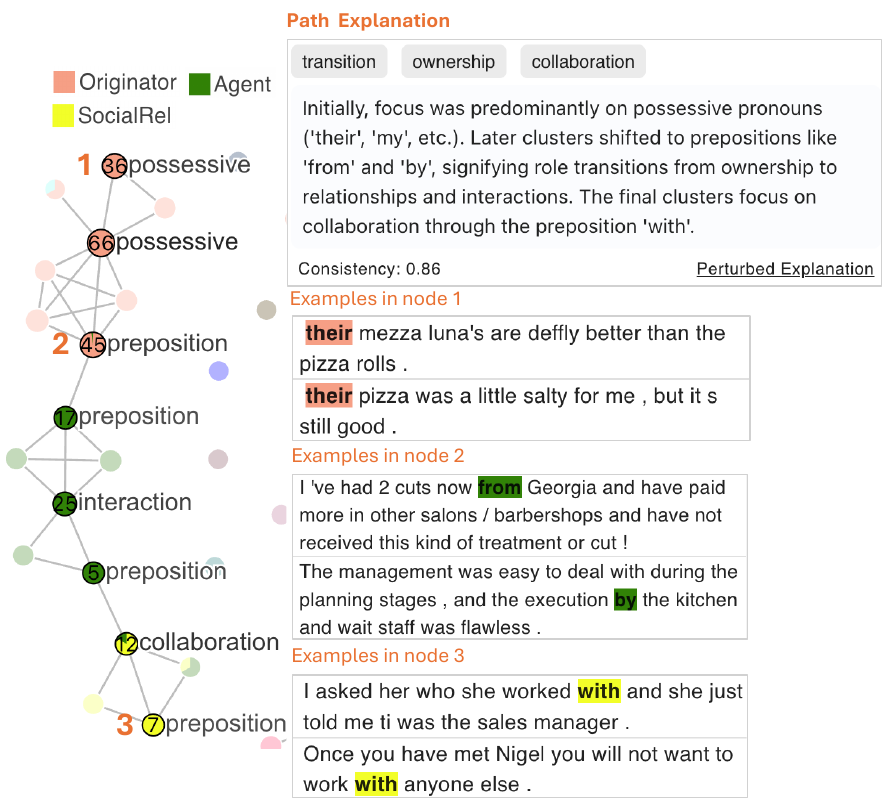}
    \vspace{-6mm}
    \caption{A path in the ball mapper.}
    \label{fig:ball-mapper-path}
    \vspace{-4mm}
\end{figure}

\para{Model Confusion.}
In addition, we identify a component mixed with three labels of ``Originator'', ``Agent'', and ``SocialRel'' as illustrated in~\cref{fig:ball-mapper-path}, suggesting that the model has difficulty in distinguishing among these categories. To investigate this, we apply the Path Explainer to analyze the shortest path connecting nodes on opposite sides of the component. The explanation indicates a progression in word usage: from possessive pronouns like \textit{`their'} and \textit{`my'}, which signal ownership, to prepositions like \textit{`from'} and \textit{`by'}, which suggest interaction, and finally to \textit{`with'}, which conveys collaboration. This linguistic transition aligns with the label shift from ``Originator'' to ``Agent'' to ``SocialRel'' along the path, indicating scenarios where the model exhibits confusion. 

\section{Groningen Meaning Bank Dataset}
\label{supp:GMB}
In the following, we use the Groningen Meaning Bank~\cite{Bos2017} dataset. This dataset provides a large collection of semantically annotated English texts with deep semantic representations. 
The dataset contains 795,935 tokens (31,175 unique tokens) and 38,405 sentences.
We use this dataset to extract layer-wise token embeddings from the pre-trained BERT-base model and generate their (classical) mapper graphs.

\begin{figure}[!ht]
    \centering
    \includegraphics[width=\linewidth]{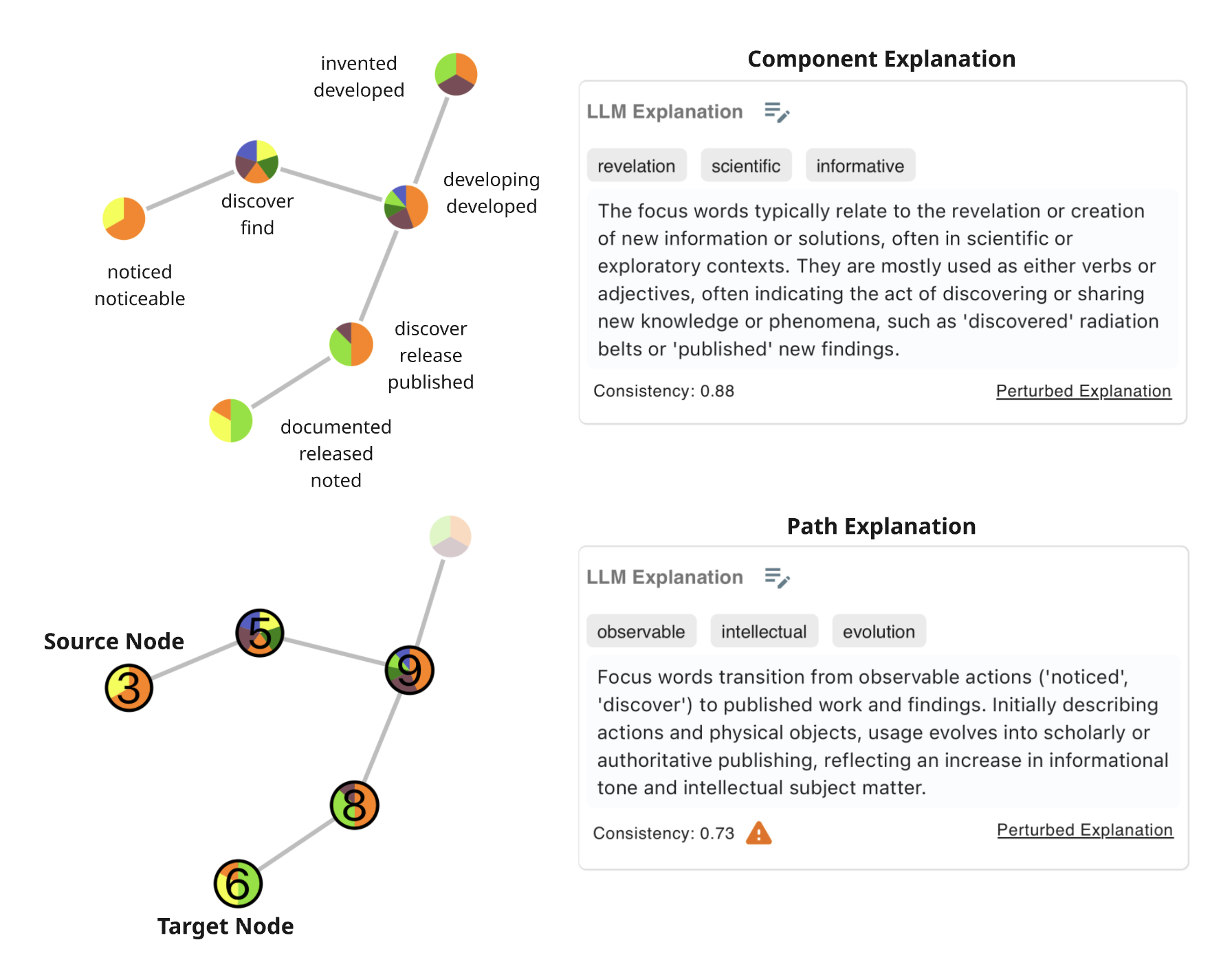}
    \vspace{-4mm}
    \caption{A component of concepts related to \textit{discovering information} in layer 1. The Component Explainer and Path Explainer are used to get insight into common characteristics among nodes and the transition between the encoded concepts along a path.}
    \label{fig:discover}
\end{figure}

\para{Synonyms Encoded in Early Layers.} 
The early layers of BERT encode the lexical properties of words and word synonymity, that is, words that have a similar semantic meaning have similar embeddings~\cite{sevastjanova2022lmfingerprints}. 
In~\cref{fig:discover}, we replicate these insights using the mapper graph. In particular, we explore a component that has mixed nodes, i.e., one cluster consists of words that have various POS tags. By utilizing the Component Explainer, it becomes apparent that the different nodes encompass words related to `\textit{awareness}', `\textit{realization}', and `\textit{development}', and have the meaning of \textit{creation of new information}. The Component Verifier produces a high consistency score for this explanation. Using the Path Explainer, we notice that focus words transition from observable actions (e.g., `\textit{noticed}') to published work and findings (e.g., `\textit{released}', `\textit{published}').

\begin{figure*}[!ht]
    \centering
    \includegraphics[width=\linewidth]{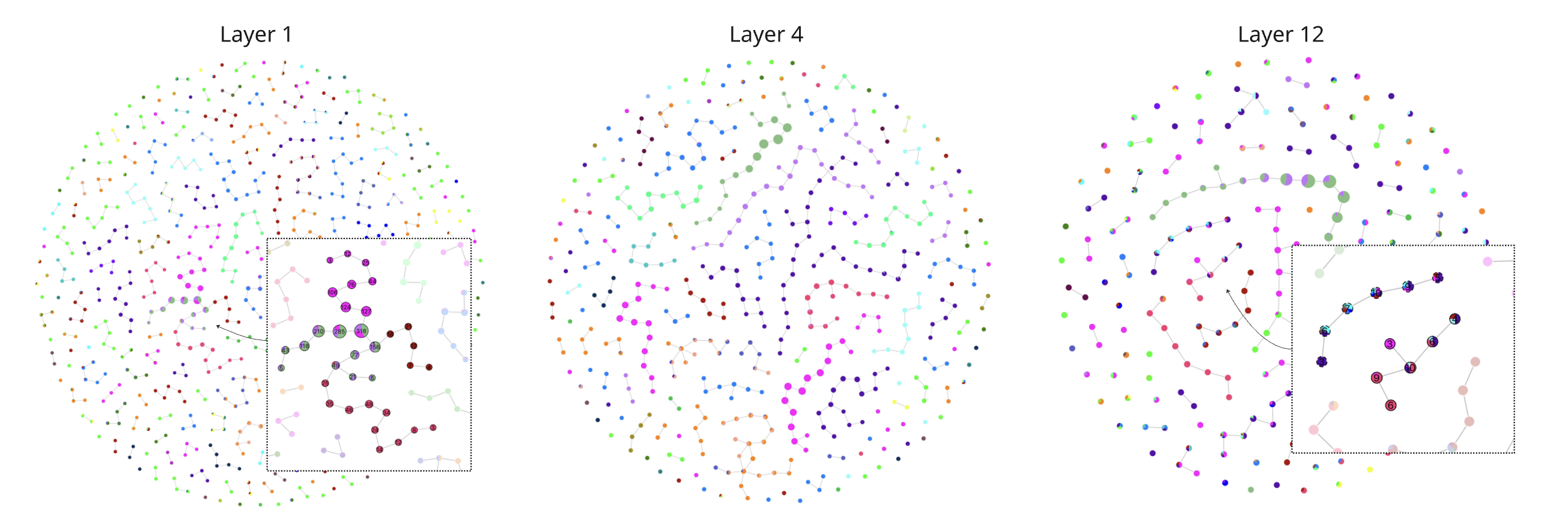}
    \vspace{-6mm}
    \caption{Layer comparison. Node colors represent token POS tags. Prior work has shown that BERT encodes POS tag information in the middle layers (e.g., layer 4). The mapper graph in layer 4 produces pure nodes, i.e., most of the components consist of words with the same POS tag.}
    \label{fig:layer-comparison}
\end{figure*}

\para{POS Tags Encoded in Middle Layers.} 
Prior work~\cite{rogers-etal-2020-primer} has shown that the middle layers of BERT encode the information on the POS tags of the words. We can replicate this insight using the mapper graph. In particular, when computing the mapper graph for different layers and representing POS tag information in nodes' color, we can observe that the purest nodes (i.e., a single color which stands for a single POS tag) are visible in the middle layers (e.g., layer 4 as shown in~\cref{fig:layer-comparison}) while other layers (e.g., layer 1 and 12) produce nodes of mixed nature.

\section{Prompt Templates and Engineering}
\label{supp:prompts}

\begin{figure}[!ht]
    \centering
    \includegraphics[width=\linewidth]{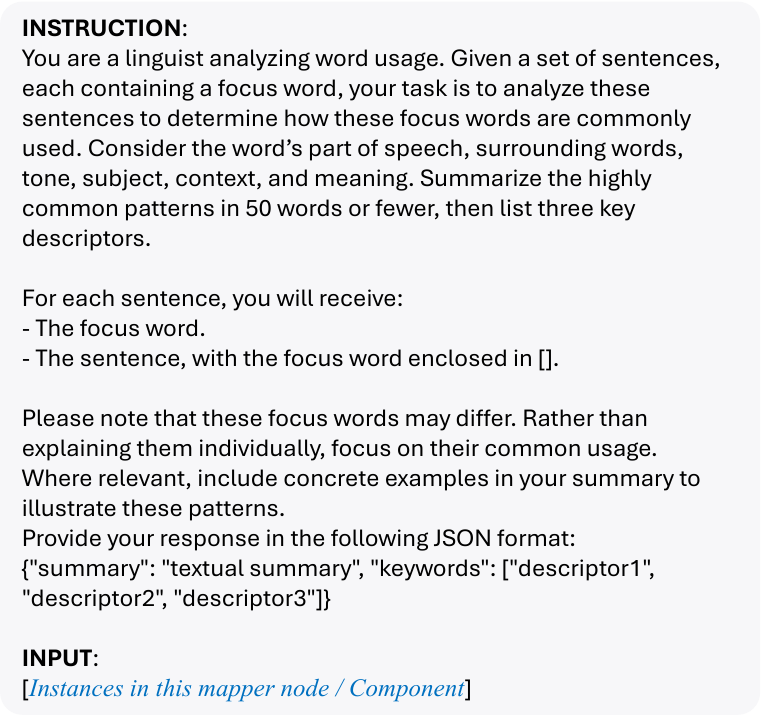}
    \caption{Prompt template for summarizing a mapper node/component.}
    \label{fig:node-explanation}
\end{figure}

\begin{figure}[!ht]
    \centering
    \includegraphics[width=\linewidth]{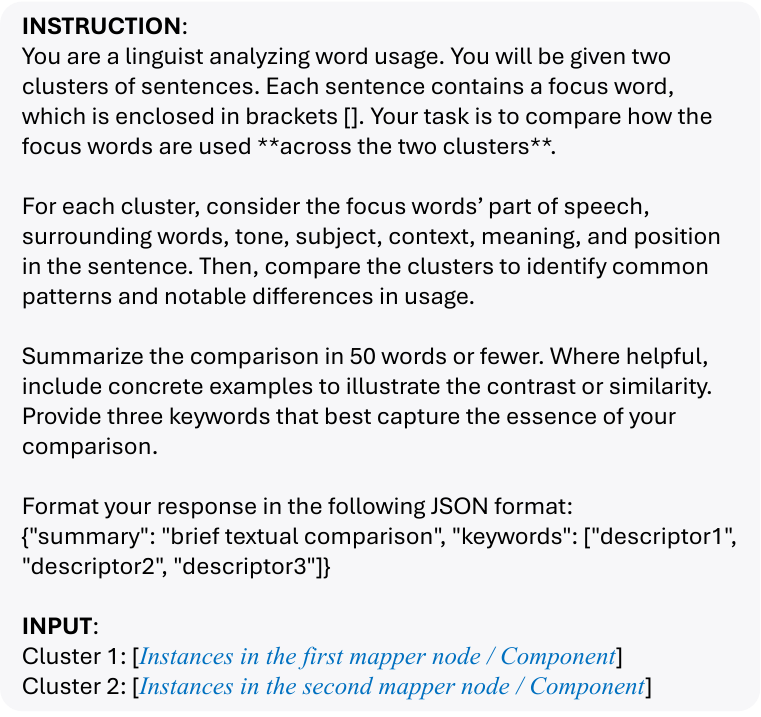}
    \caption{Prompt template for comparing two mapper nodes/components.}
    \label{fig:node-compare}
\end{figure}

\begin{figure}[!ht]
    \centering
    \includegraphics[width=\linewidth]{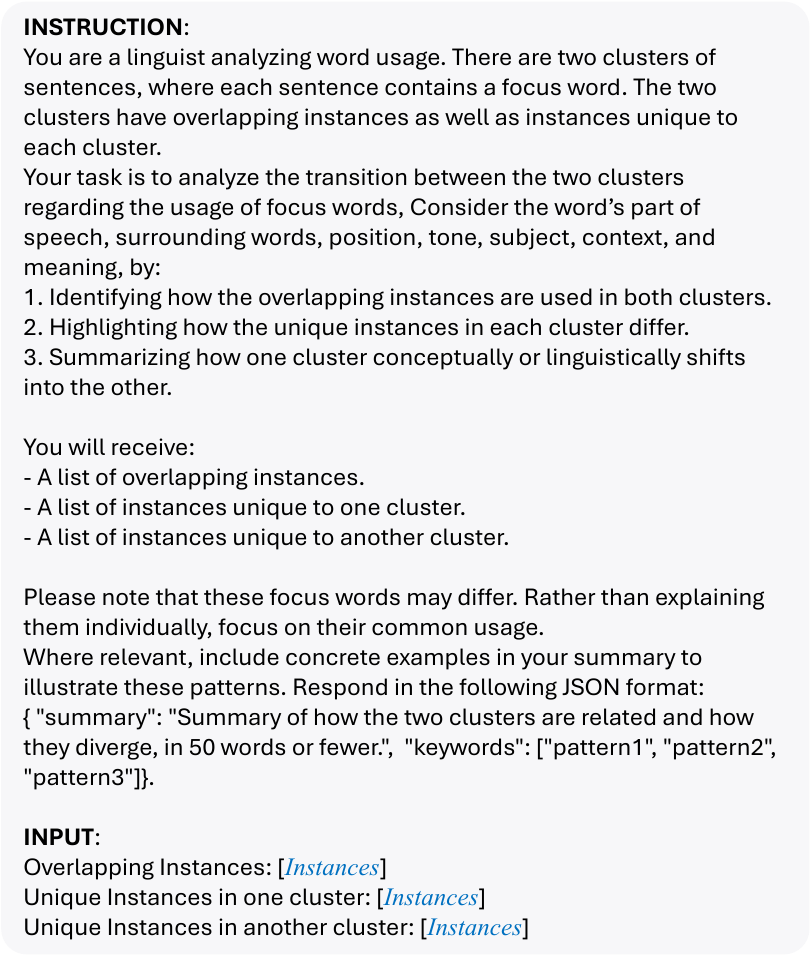}
    \caption{Prompt template for summarizing a mapper edge.}
    \label{fig:edge-explanation}
\end{figure}

\begin{figure}[!ht]
    \centering
    \includegraphics[width=\linewidth]{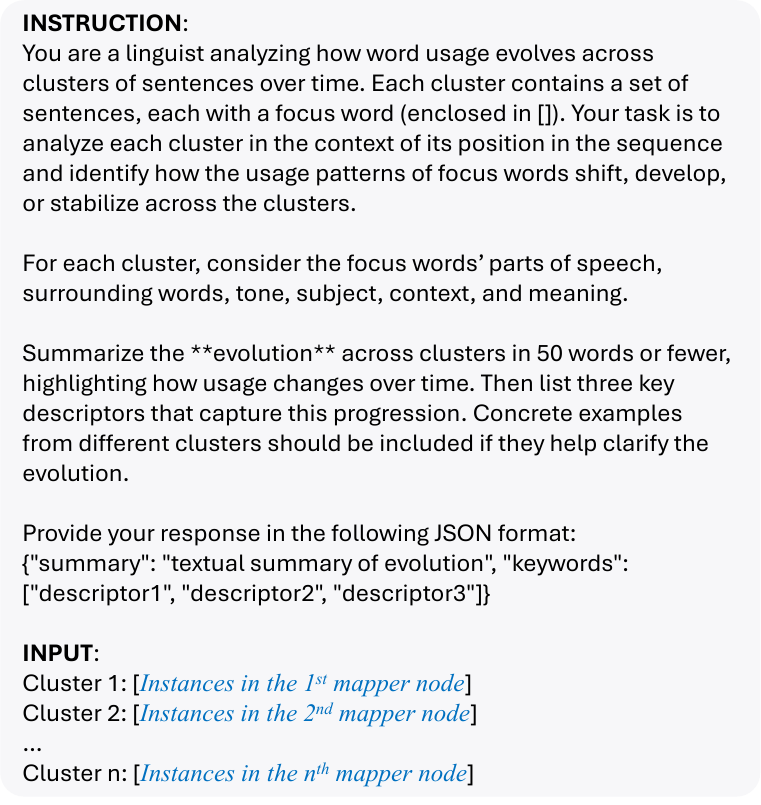}
    \caption{Prompt template for summarizing a mapper path.}
    \label{fig:path-explanation}
\end{figure}

\begin{figure}[!ht]
    \centering
    \includegraphics[width=\linewidth]{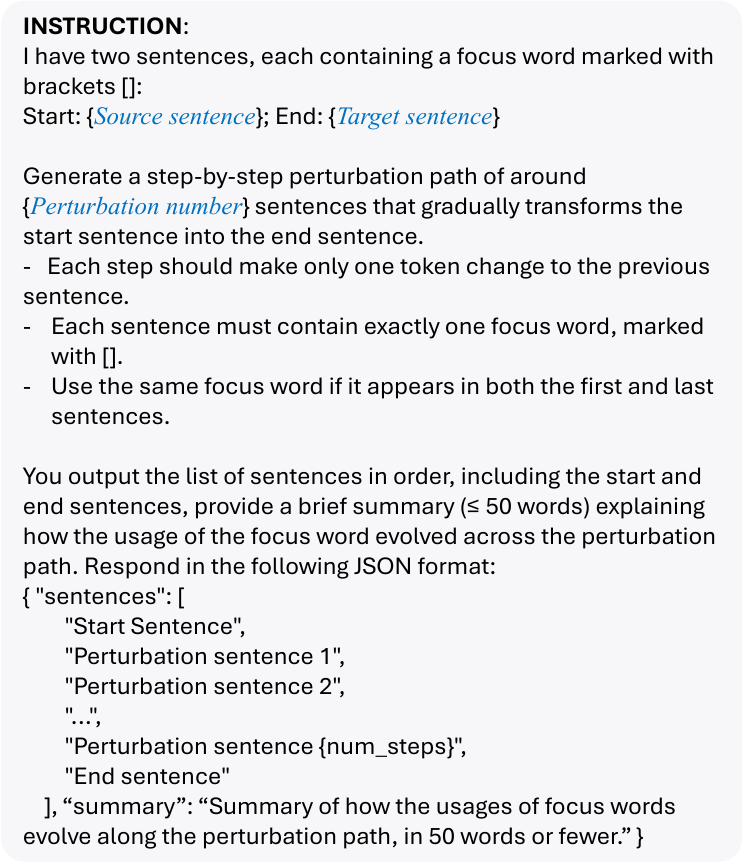}
    \caption{Prompt template for creating a perturbation trajectory between two sentences.}
    \label{fig:trajectory-prompt}
\end{figure}

\begin{figure}[!ht]
    \centering
    \includegraphics[width=\linewidth]{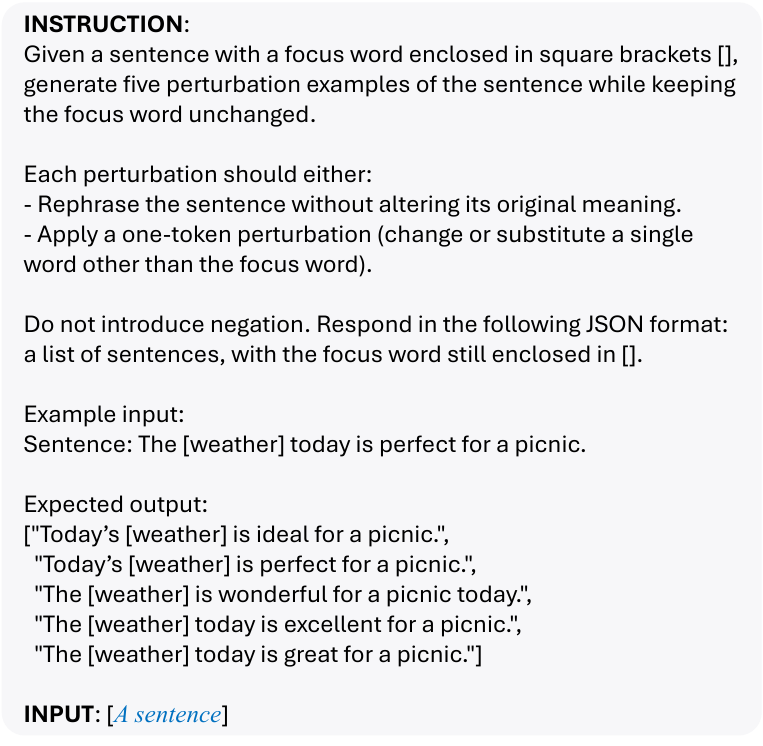}
    \caption{Prompt template for generating the perturbed sentences of a given sentence.}
    \label{fig:perturbation-prompt}
\end{figure}

\para{Prompt Templates.} For the Node and Component Explainers, \cref{fig:node-explanation} and~\cref{fig:node-compare} present the prompt templates used to summarize a  node/component and to compare two nodes/components. 
For the Edge Explainer, \cref{fig:edge-explanation} shows the prompt template to summarize an edge. For the Path Explainer, \cref{fig:path-explanation} show the prompt templates for summarizing a path.
For the Trajectory Explainer, \cref{fig:trajectory-prompt} shows the prompt template used to generate a sequence of 1-token perturbations, given the source and target sentences along with an approximate trajectory length. 
In addition, \cref{fig:perturbation-prompt} presents the prompt template used to generate perturbed sentences of a given sentence through 1-token replacement and rephrasing. This template is used by the Node, Edge, Path, and Component Verifier.

\begin{figure*}[!ht]
    \centering
    \includegraphics[width=\linewidth]{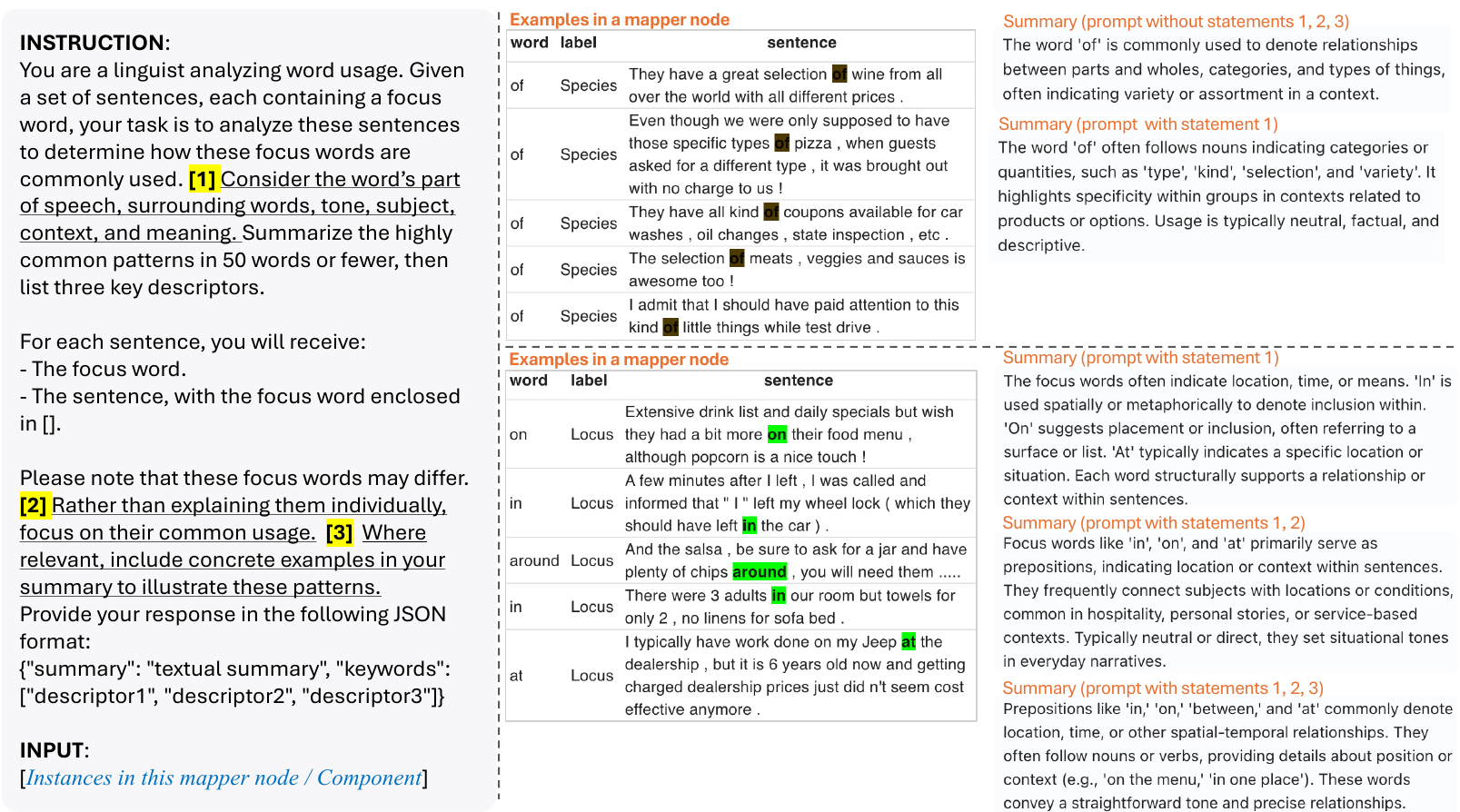}
    \caption{An example of prompt engineering for the mapper node summary. Left: the final prompt, with Statements 1–3 added incrementally. Top right: node summaries generated with and without Statement 1. Bottom right: node summaries generated as Statements 2 and 3 are gradually added.}
    \label{fig:prompt-engineer}
\end{figure*}

\para{Prompt Engineering.} To fully unleash the capabilities of LLMs, we employ established prompt engineering techniques in our initial prompts, such as role-playing and explicit input-output definitions. 
We then iteratively refine these prompts through qualitative experimentation, initially focusing on the mapper node summary. 
Once validated, the refined strategies are extended to prompts used in other explanation tasks.
As shown in~\cref{fig:prompt-engineer}, the node summary prompt undergoes three main iterations, each introducing an additional guiding statement (1–3).
For example, Statement 1 explicitly highlights linguistic properties, encouraging the LLM to produce more linguistically grounded responses.
Statement 2 prompts LLMs to consider a cluster of words as a whole rather than treating them based on lexicon; Statement 3 prompts LLMs to provide concrete examples to support its output, making the explanation more comprehensible and relatable.

\end{document}